\documentclass[fleqn,usenatbib]{mnras}

\usepackage{newtxtext,newtxmath}

\usepackage[T1]{fontenc}
\usepackage{ae,aecompl}

\usepackage{lineno}
\modulolinenumbers[5]

\usepackage{hyperref}
\usepackage{amsmath}
\usepackage{amssymb}

\usepackage{etoolbox}
\makeatletter
\patchcmd\@combinedblfloats{\box\@outputbox}{\unvbox\@outputbox}{}{\errmessage{\noexpand patch failed}}
\makeatother

\usepackage{psfrag} 

\usepackage{media9} 

\usepackage[utf8]{inputenc} 
\usepackage{graphicx}

\usepackage{relsize}
\usepackage{pbox}
\usepackage[usenames,dvipsnames]{color}
\definecolor{aliceblue}{rgb}{0.94, 0.97, 1.0}
\definecolor{lightgray}{rgb}{0.83, 0.83, 0.83}
\usepackage{bm}

\usepackage{array}
\usepackage{ulem}

\newcommand{\jcop}{J. Comput. Phys.}

\newcommand{\Pc}{P_{\rm c}}
\newcommand{\BB}{\tilde{B}}
\newcommand{\EE}{\tilde{E}}
\newcommand{\JJ}{\tilde{J}}
\newcommand{\TT}{\tilde{T}}
\newcommand{\Rstar}{13.7\,\text{km}}
\newcommand{\Rstars}{1.37\times 10^{6}\,\text{cm}}
\newcommand{\Lo}{2.8\times 10^{49}\,\text{erg}\,\text{s}^{-1}}
\newcommand{\etao}{110}
\newcommand{\etaaster}{393}
\newcommand{\rrho}{\tilde{\rho}_{\rm e}}
\newcommand{\ee}{\mathcal{E}}

\newcommand{\vect}[1]{\boldsymbol{#1}}

\title[Instability of twisted magnetar magnetospheres]{Instability of twisted magnetar magnetospheres}

\author[J. F. Mahlmann, T. Akgun, J. A. Pons, M. A. Aloy and P. Cerdá-Durán]
{J. F. Mahlmann$^1$\thanks{jens.mahlmann@uv.es}, T. Akgün$^2$, J. A. Pons$^2$, M.A. Aloy$^1$ and P. Cerdá-Durán$^1$\thanks{pablo.cerda@uv.es}\\ $^1$Departament d'Astronomia i Astrofísica, Universitat de València, 46100, Burjassot, Spain\\
 $^2$Departament de Fisica Aplicada, Universitat d'Alacant, 03690, Alicante, Spain}

\date{Accepted 2019 September 25. Received 2019 September 22; in original form 2019 July 31}

\pagerange{\pageref{firstpage}--\pageref{lastpage}} \pubyear{2019}

\begin{document}

\maketitle

\label{firstpage}

\begin{abstract}
 We present three-dimensional force-free electrodynamics simulations of magnetar magnetospheres that demonstrate the instability of certain degenerate, high energy equilibrium solutions of the Grad-Shafranov equation. This result indicates the existence of an unstable branch of twisted magnetospheric solutions and allows to formulate an instability criterion. The rearrangement of magnetic field lines as a consequence of this instability triggers the dissipation of up to 30\% of the magnetospheric energy on a thin layer above the magnetar surface. During this process, we predict an increase of the mechanical stresses onto the stellar crust, which can potentially result in a global mechanical failure of a significant fraction of it. We find that the estimated energy release and the emission properties are compatible with the observed giant flare events. The newly identified instability is a candidate for recurrent energy dissipation, which could explain part of the phenomenology observed in magnetars.
\end{abstract} 


\section{Introduction}
\label{sec:introduction}

Soft gamma-ray repeaters (SGRs) are neutron stars with recurrent X-ray activity in the form of short bursts with duration $\sim 0.1$\,s and luminosities in the range $10^{36}-10^{43}$\,erg\,s$^{-1}$. Over the last 40 years, three bursts have been uniquely energetic, the so-called {\it giant flares} (GFs) with luminosities of the order of $10^{44}-10^{47}$\,erg\,s$^{-1}$ \citep[SGR\,0525-66, SGR\,1900+14 and SGR\,1806-20, see][]{Cline1980,Hurley1999,Hurley2005}. In the three referenced cases, a short initial peak was followed by a softer X-ray tail lasting for $50-400$\,s. The engine behind these extraordinary events are {\it magnetars}, neutron stars with the strongest known magnetic fields \citep[$10^{14}-10^{16}$\,G; see comprehensive reviews of magnetar observations and physics, e.g. in][]{Woods2006,Rea2011,Turolla2015,Mereghetti2015,Kaspi2017}.

The precise mechanism producing such energetic events is still unclear. Strong magnetic fields are a gigantic energy reservoir in magnetars, generally of the order 
\begin{align}
\ee_{\rm magnetar} \sim 1.6\times 10^{47} \,\text{erg}\,\left ( \frac{B}{10^{15}\,\text{G}}\right )^{2}\left ( \frac{R_*}{10\,\text{km}}\right )^3,
\end{align}
where we consider a neutron star with radius $R_*$.

The timescale on which the magnetar is evolving, mainly due to Hall drift and Ohmic dissipation in the crust, is of the order of $10^3-10^6$\,yrs \citep{Jones1988,Goldreich1992,Pons2007,Pons2009,Gourgouliatos2016}, by itself too slow to explain this phenomenology. Two complementary models have tried to explain these observations. In the {\it crustquake} model \citep{Thompson1996,Perna2011} the dynamical trigger is the mechanical failure of patches of the magnetar crust due to large stresses built during its magneto-thermal evolution. Numerical simulations of the Hall evolution of the crust \citep{Vigano2013} show that it is possible to recurrently reach the maximum stress supported by the very same \citep{Horowitz2009,Baiko2018}. At this point, the crust likely becomes plastic \citep{Levin2012}, i.e. the crust generates thermo-plastic waves emerging from such a localized trigger, or in other words \textit{yields} \citep{Beloborodov2014,Li2016}. The waves propagate into the magnetosphere, probably resulting in rapid dissipation through a turbulent cascade triggered by reconnection on slightly displaced flux surfaces \citep{Thompson1996,Thompson2001,Li2018}. The energy released in those events suffices to explain the observed luminosities, even for GFs \citep{Thompson1996,Lander2015}. The burst duration ($\sim 0.1$\,s) is related to the crossing time of shear waves through the whole crust ($1-100$~ms). A limitation is that, if stressed for long periods of time ($\sim 1\,$yr) as it is the case due to the slow magneto-thermal evolution, the crust may yield at significantly lower breaking stresses \citep{Chugonov2010}. In that case, it would effectively deform as a plastic flow, and, depending on its (unknown) properties, cease to yield altogether \citep{Lyutikov2015,Lander2019}. \citet{Thompson2017} has argued that even in this case the crust could yield.

The {\it magnetospheric instability} model requires a strongly twisted magnetosphere that becomes unstable and leads 
to a rapid reconnection event \citep{Lyutikov2003}. The existence of long-lived magnetospheric twists is supported by the observation
of hard X-ray emission in persistent magnetars \citep{Beloborodov2013b,Hascoet2014}. During the magneto-thermal evolution of the crust,
the displacement of the magnetic field footprints can generate large twists in the magnetosphere \citep{Akgun2017,Akgun2018b}. 
Above a critical twist, the magnetosphere becomes unstable and undergoes a rapid rearrangement where energy is dissipated by 
reconnection \citep{Lyutikov2003,Gill2010,Elenbaas2016} in a similar fashion as in the crustquake model. The main challenge of this scenario
is the ability of the crust to produce significant twists in the magnetosphere. \citet{Beloborodov2009} estimated that currents supporting 
magnetospheric twist are bound to dissipate on timescales of years, effectively leading to a progressive untwisting. Therefore, Hall evolution is required to proceed relatively fast in order to allow for significant twists. Plastic viscosity may also be a problem for similar reasons \citep{Lander2019}. The latter authors have also suggested that the dynamical crust
fractures of the crustquake model could be substituted by sustained episodes of accelerated plastic flows which are able to generate large magnetospheric twists on times shorter than the untwisting timescale. 

Numerical simulations by \citet{Parfrey2012,Parfrey2013} and \citet{Carrasco2019} confirm the instability of the magnetosphere beyond a critical twist, accompanied by the formation of plasmoids. These results are an analogy to the context of eruption processes in the solar corona as found in numerical experiments
by \citet{Roumeliotis1994,Mikic1994}. The energy dissipated in the reconnection events is sufficient to explain the GF processes \citep{Parfrey2012}. A caveat to these simulations is that the applied twisting rate is larger than the one expected from the respective magneto-thermal evolution, although it would be fine if the trigger was a rapid plastic deformation. 

An alternative approach to the above is the study of stability properties of magnetospheres. A number of authors have constructed equilibrium solutions to the Grad-Shafranov equation (GSE) for neutron star magnetospheres \citep{Glampedakis2014,Fujisawa2014,Pili2015,Akgun2016,Kojima2017,Kojima2018,Kojima2018b,Akgun2018a}. \citet{Akgun2017} performed magneto-thermal evolutions coupling the crustal magnetic field at the stellar surface with an exterior equilibrium solution. The results showed that large twists grow in the magnetosphere up to a critical point beyond which no stable equilibrium solutions where found. A more detailed analysis by \citet{Akgun2018a} showed that, for sufficiently large twists, the solutions of the GSE are degenerate with several possible configurations of different energies but matching boundary conditions at the surface. This suggests the possibility of an unstable branch of the solutions and, thus, a possible explanation for the occurrence of bursts and GFs. In this work we explore this possibility by performing three-dimensional (3D) numerical simulations of the equilibrium models in \citet{Akgun2018a}. We asses their stability properties and their potential as candidates for transient magnetar phenomenology.

This work is organised as follows. In section~\ref{sec:physics} we review and discuss the physics involved in magnetars relevant to the processes that we want to study. In section~\ref{sec:ff_equations} we briefly review the equations of force-free electrodynamics (FFE) implemented for simulations conducted on the infrastructure of the \texttt{Einstein Toolkit} (supplemented by appendix \ref{sec:augmented_system}). A detailed description of the derivation of initial models according to \citet{Akgun2018a} is given in section~\ref{sec:twisted_models}. In section~\ref{sec:simulations} we present the numerical setup of our simulations as well as the outcome of the conducted 3D force-free simulations of twisted magnetospheres (reviewing details on maintaining the force-free regime in appendix \ref{sec:ff_breakdown}). The observed rapid dissipation of electromagnetic energy through the magnetar crust is interpreted and related to observable quantities, such as luminosity estimates, shear stresses on the stellar crust, and opacity models, in section~\ref{sec:discussion}. Along this paper we use Gaussian units in CGS, except for section~\ref{sec:ff_equations} in which we use Heaviside-Lorentz with geometrised units ($G=c=M_\odot=1$). For convenience we express current densities in A m$^{-2}$ and voltages in V, instead of the corresponding CGS units.

\section{Physics of magnetars}
\label{sec:physics}
 
 The basic structure of the magnetar interior is a (likely) fluid core of $\sim 10$~km radius, amounting for most of the mass of the object, surrounded by a solid crust of about $1$~km size.
Outside, there is a tenuous, co-rotating magnetosphere connected to the NS by magnetic field lines (threading the central object) that extend up to the light cylinder at distances larger than $10^5$~km. We start by discussing 
some basic properties of the different parts of the magnetosphere relevant for the interpretations and models presented later in this work.

\subsection{Currents supporting the magnetosphere}
\label{sec:current_magnetsophere} 

For the typical rotation periods of magnetars ($P\sim 1$ - $10$\,s) the Goldreich-Julian particle density \citep{GJ1969} for a magnetar magnetosphere has the typical value
\begin{align}
n_{\textsc{gj}} = 7\times 10^{12} \,\text{cm}^{-3}\,\left ( \frac{B_{\rm pole}}{10^{15} \,{\rm G}}\right) \left( \frac{R_{*}}{r}\right )^3 \left ( \frac{10 \,{\rm s}}{P}\right),
\end{align}
where $B_{\rm pole}$ is the magnetic field strength at the magnetar pole, $R_*$ the magnetar radius and $r$ the distance to the center of the star. This limits the magnetospheric current density close to the surface to $J<e\,c\,n_{\textsc{gj}} \approx 3\times 10^8$\,A/m$^2$, much below the typical values needed to support currents in strongly twisted magnetospheres of magnetars, of the order of 
\begin{equation}
 J\sim \frac{B c}{4 \pi r}\sim 8.2\times 10^{12}\,\text{A}\,\text{m}^{-2}\,\left( \frac{B_{\rm pole}}{10^{15}\,\text{G}}\right ) \left( \frac{R_*}{10\,\text{km}}\right )^{-1}\label{eq:GJcurrent}.
\end{equation}
In general, magnetospheric currents in magnetars cannot be supported neither by Goldreich-Julian charges
nor by charges lifted from the surface. \citet{Beloborodov2007} proposed that the currents are supported by $e^+$-$e^-$ pairs
generated in the magnetosphere in an intermittent discharge process that can be sustained for voltages along magnetic field
lines of about $10^8 -10^9$\,V. This voltage can be maintained by self-induction in untwisting magnetospheres \citep{Beloborodov2009}.
This untwisting is driven by the effective resistivity of the magnetosphere;
the thermal photons from the magnetar's surface scatter resonantly off the charges supporting the magnetospheric currents, taking energy away, 
at the same time that pairs are produced. The untwisting timescale is $\sim 1$\,yr, and it may explain the spectral evolution of some magnetars 
\citep{Beloborodov2009}.

\subsection{Timescales}
\label{sec:timecales} 

Changes in magnetars take place during two different timescales. On the one hand, there is a {\it secular timescale} of thousands of years during which the magnetar is essentially in equilibrium. On the other hand, there is a {\it dynamical timescale} associated to energetic events (burst, flares) that can produce observable variations on timescales as fast as $0.1$\,s. The latter are likely associated to out-of-equilbrium states.

\subsubsection{Secular timescales}
\label{sec:secular}

The {\it secular timescale} is set by the slow magnetothermal evolution of the cooling object. The interior magnetic field evolution is dominated by Hall drift and Ohmic diffusion at the crust \citep[see, e.g.][and references therein]{Vigano2012,Fujisawa2014}, which proceeds on typical timescales of $10^3-10^6$~yr. The long-term evolution of the magnetosphere is driven by the changes in the crustal magnetic field, which displaces the footprints of the magnetospheric magnetic field lines. Since this evolution is much slower than the dynamical timescale of the magnetosphere (see below),
it can be considered that the magnetosphere evolves through a series of equilibrium states. This evolution creates a twist in the magnetosphere supported by currents - until a critical maximum twist is reached ($\varphi_{\rm crit} \sim 1$\,rad) beyond which no magnetospheric equilibrium solutions exist \citep{Akgun2017}. The stability of the magnetosphere close to this critical point is the subject of this paper.

At the same time as the crustal magnetic field evolves, other processes in the magnetosphere can also contribute to the evolution. The untwisting of the magnetosphere
on timescales of $\sim1$\,yr \citep[][and discussion in section~\ref{sec:secular}]{Beloborodov2009}, may be a competing action to the twisting process described above.

Although the velocity of the footprints is typically very slow, numerical simulations of the magnetothermal evolution of magnetars including the magnetosphere show that, close to the critical point, it can be as fast as $v_{\varphi}\sim 1$\,km\,yr$^{-1}$ \citep[see][]{Akgun2017} in the most optimistic scenario. Therefore, close to the critical twist, the magnetosphere twists slowly ($\dot\varphi_{\rm max,crit} \,\lesssim\, 0.1$\,rad\,yr$^{-1}$), evolving on timescales $\,\gtrsim\, 10$\,yrs. In the best case scenario, this timescale is comparable to the untwisting timescale ($\sim1$\,yr) and, hence, parts of the magnetosphere could sustain a significant twist. This timescale is still much longer than the dynamical timescale of the system (see below). Therefore, in our study of the dynamical behavior we can neglect the secular evolution of the field.

\subsubsection{Dynamical timescales}
\label{sec:dynamical}

The {\it dynamical timescale} is set by the travel time of waves propagating in the different regions of the magnetar. In the magnetosphere, the mass density can be neglected in view of the dominating magnetic field energy density. Also, the velocity of Alfvén and fast magnetosonic waves is degenerated to the speed of light. Hence, within $\sim 100$\,km the whole magnetosphere is coupled through timescales smaller than $1$\,ms, which sets the scale for the dynamical evolution of the magnetosphere. In this region it is possible to neglect the inertia of the fluid in the evolution equations of so-called FFE, which is used in the numerical simulations of this work.

In the outermost parts of the crust, the force-free condition still holds because of low densities. At sufficiently
high densities, elastic forces of the solid crust and pressure gradients break this condition. To estimate the transition density
one may consider the depth at which waves propagate at a velocity significantly different to the speed of light. Two possible waves can travel in the interior of the magnetised crust, the so-called magnetosonic (ms) waves, related to sound waves, and 
magneto-elastic (me) waves, a combination of Alfv\'en and shear waves. The complete eigenvalue structure of relativistic ideal MHD equations
in the presence of an elastic solid is not known. To make a simple order of magnitude estimate of the different wave speeds, we use
the expression of magneto-elastic torsional waves parallel to the magnetic field derived in \cite{Gabler2012} as well as the expression
for fast magnetosonic waves perpendicular to the field\footnote{Slow magnetosonic waves are also possible 
but their velocity is much smaller and not relevant for this work, in fact, for the case of waves perpendicular to the magnetic field their speed 
is zero.}:
\begin{align}
v_{\rm me} /c = \sqrt{\frac{\mu_s+B^2}{e + B^2}} \qquad \qquad v_{\rm ms} /c= \sqrt{\frac{e c_s^2 + B^2}{e+B^2}},
\end{align}
where $e$ is the energy density and $\mu_s$ the shear modulus. Note that in the limit of low magnetic field
($B^2\ll \mu_s, \, B^2\ll e$) we recover the shear and sound speed, respectively. In the high magnetic field limit 
($B^2\gg \mu_s,\, B^2\gg e$) both, $v_{\rm me}$ and $v_{\rm ms}$, coincide with the speed of light. Inside the fluid core ($\mu_s=0$) the magneto-elastic speed
becomes the Alfv\'en speed. 

\begin{figure}
	\centering
	\includegraphics[width=0.5\textwidth]{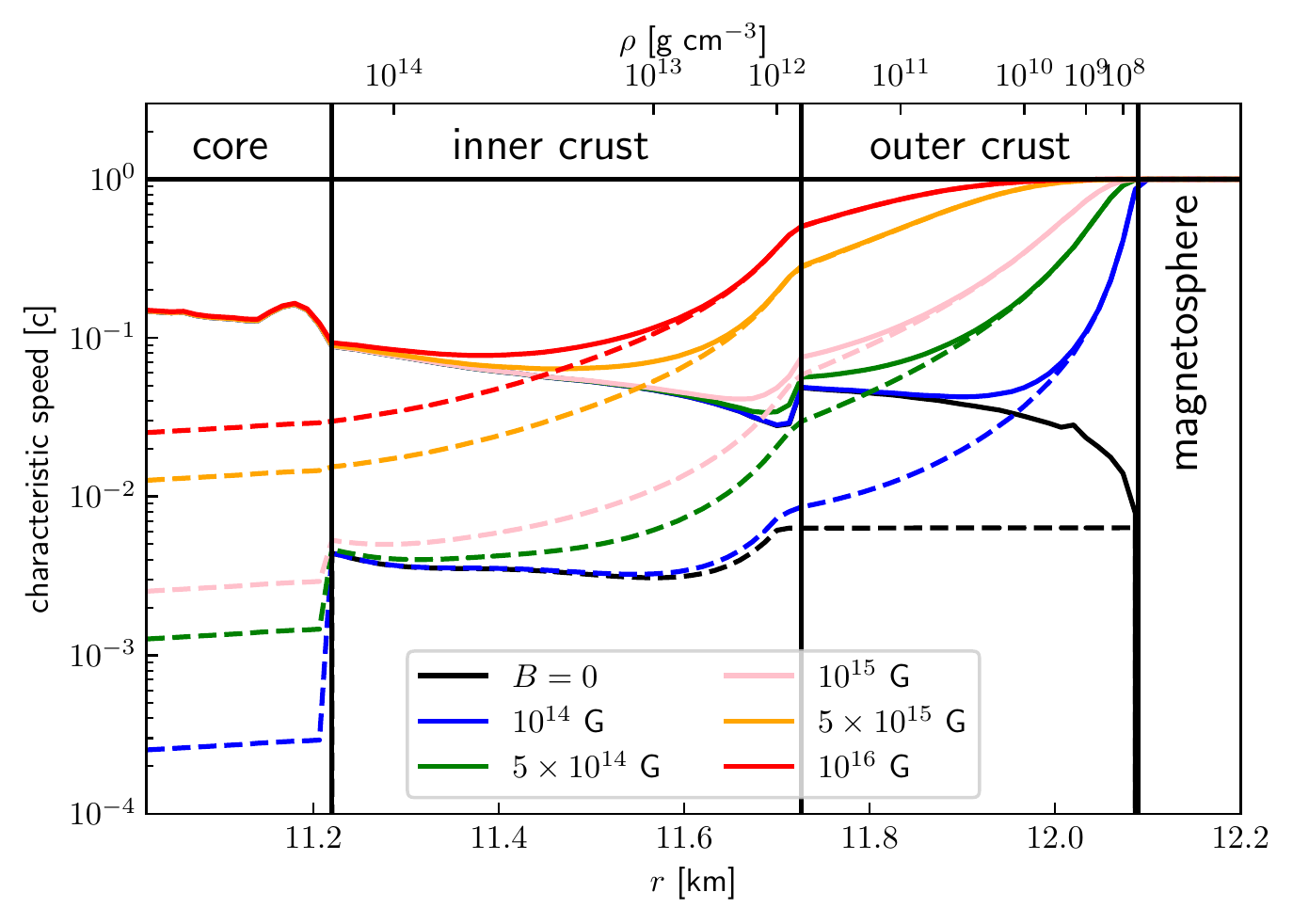} 
	\caption{Fast magnetosonic (solid lines) and magnetoelastic (dashed lines) speed in the outer layers of a magnetar, for different 
	magnetic field strengths ranging from 0 to $10^{16}$\,G. The neutron star model corresponds to the $1.4 M_\odot$ mass APR+DH model of \citet{Gabler2012}.
	The magnetic field is considered to be constant for simplicity. }
	\label{fig:speeds}
\end{figure}

Figure~\ref{fig:speeds} shows the value of the characteristic speeds in the outer layers 
of a typical NS model for different magnetic fields in the magnetar range. Indeed, both fast magnetosonic waves and Alfvén waves have a degenerate speed equal to the speed of light in the magnetosphere.
Inside the outer crust ($\rho < 4\times 10^{11}$\,g\,cm$^{-3}$), all characteristic speeds transition from the speed of light to a significantly lower value,
in a region that can still be considered force-free. This transition depends on the magnetic field
strength, happening deeper inside the star for larger values of $B_{\rm pole}$. Given these characteristic speeds, any global rearrangement 
of the magnetosphere can modify the entire structure of the crust (of size $\sim 2 \pi R_*$) on a timescale of $\sim 1$~ms for magnetosonic 
waves and $\sim 10$\,ms for magnetoelastic waves.

One last aspect to consider is the ability of magnetospheric waves to transmit energy into the crust. The discussion should be limited to Alfv\'en waves, which become magnetoelastic waves once they penetrate the crust; the energy carried by fast magnetosonic waves in the magnetosphere can be neglected due to the small density, which renders the compressibility effects of fast-magnetosonic waves unimportant.

Since the characteristic time in the magnetosphere is $\sim 1$\,ms, the typical frequency of the waves generated during
its dynamics is in the kHz range. At this frequency, the crust can be considered as a thin layer because its thickness ($\sim 1$\,km) is much smaller
than the typical wavelength in the magnetosphere ($\lambda \sim 100\,$km). In this case the energy transmission coefficient for waves perpendicular 
to the surface is approximately \citep[cf.][]{Link2014,Li2015}
\begin{equation}
\mathcal{T} =\frac{4 v_{\rm me}/c}{(1+v_{\rm me}/c )^2}\approx 0.04 \left (\frac{v_{\rm me}/c}{0.01} \right), 
\label{eq:transmission}
\end{equation}
for typical physical conditions in the magnetar crust. Given the low transmission coefficients of magnetospheric Alfvén waves hitting the crust as well as the differences on timescales between the crust and the magnetosphere (typically $\sim 10$ times shorter in the later) it is reasonable to consider that most of the crust remains rigid during any dynamical rearrangement of the magnetosphere.

In our magnetar model we will consider two regions: A force-free region ({\it exterior}, hereafter) consisting of the magnetosphere and the force-free part
of the outer crust as well as the magnetar {\it interior} for the remainder of the NS, which we will consider to be fixed during our simulations. The 
limit between both regions is a spherical surface below the NS surface, where magnetic field lines are anchored, and is located below the transition density between inner and outer crust at a density 
$\rho<4\times 10^{11}\,$g\,cm$^{-3}$. For the purpose of describing the simulations we 
will refer to this transition point simply as \textit{surface}.

%

\section{Force-free Electrodynamics}
\label{sec:ff_equations}

In analogy to \citet{Komissarov2004} and \citet{Parfrey2017} we solve Maxwell's equations in the force-free limit:
\begin{align}
\frac{\partial \vect{\BB}}{\partial t} = -\vect{\nabla}\times\vect{\EE}\qquad\text{and}\qquad\frac{\partial \vect{\EE}}{\partial t} =\vect{\nabla}\times\vect{\BB} - \vect{\JJ}_{\rm \text{\tiny FF}}\:,
\label{eq:maxwell}
\end{align}
where $\vect{\EE}$, $\vect{\BB}$, and $\vect{\JJ}_{\rm \text{\tiny FF}}$ are the electric field, the magnetic field, and the so-called force-free current, respectively. We place a tilde to distinguish quantities expressed in our Heaviside-Lorentz \textit{geometrised} (HLG) system of units, while the same symbols without tilde express quantities in the Gaussian \textit{non-geometrised} (GNG) system of units (see Table~\ref{tab:unit_conversions}). We explicitly include the charge conservation equation 
\begin{align}
 \frac{\partial \rrho}{\partial t}+\vect{\nabla}\cdot\vect{\JJ}_{\rm \text{\tiny FF}} = 0\,,
 \label{eq:chargecons}
\end{align}
where $\rrho$ represents the charge density. Furthermore, we use a mixed hyperbolic/parabolic correction by the introduction of additional potentials (further discussed in appendix \ref{sec:augmented_system}) in order to numerically ensure the constraints $\vect{\nabla}\cdot\vect{\BB}=0$ and $\vect{\nabla}\cdot\vect{\EE}=\rrho$ \citep{Dedner2002,Palenzuela2009,Mignone2010}.

In the force-free limit it is necessary to guarantee that there are either no forces acting on the system or, more generally, that the forces of the system balance each other. This is equivalent to a vanishing net Lorentz force on the charges $\rrho$ \citep[see, e.g.][]{Camenzind2007}:
\begin{align}
\vect{\EE}\cdot\vect{\JJ}_{\rm \text{\tiny FF}}&=0\label{eq:electromagnetic_work}\\
\rrho\vect{\EE}+\vect{\JJ}_{\rm \text{\tiny FF}}\times\vect{\BB}&=0\label{eq_lorentz_force}
\end{align} 
From equation~(\ref{eq_lorentz_force}) one readily obtains the degeneracy condition
\begin{align}
\vect{\EE}\cdot\vect{\BB}&=0.\label{eq:force_free_crossfield}
\end{align}
Additionally, force-free fields are required to be magnetically dominant, the magnetic field being always stronger than the electric one, such that the following condition must hold:
\begin{align}
\vect{\BB}^2-\vect{\EE}^2&\geq 0.\label{eq:force_free_dominance}
\end{align}
Conditions (\ref{eq:force_free_crossfield}) and (\ref{eq:force_free_dominance}), as well as the conservation condition $\partial_t\left(\vect{\EE}\cdot\vect{\BB}\right)=0$ can be combined in order to obtain an explicit expression for $\vect{\JJ}_{\rm \text{\tiny FF}}$ \citep[cf.][]{Komissarov2011,Parfrey2017}:
\begin{align}
\vect{\JJ}_{\rm \text{\tiny FF}}=\left[\vect{\BB}\cdot\vect{\nabla}\times\vect{\BB}-\vect{\EE}\cdot\vect{\nabla}\times\vect{\EE}\right]\frac{\vect{\BB}}{\BB^2}+\rrho\frac{\vect{\EE}\times\vect{\BB}}{\BB^2}\label{eq:ff_current}
\end{align}
Across the literature \citep[e.g.][]{Komissarov2004,Alic2012,Parfrey2017} we find various modifications in the definition of $\vect{\JJ}_{\rm \text{\tiny FF}}$ in order to drive the numerical solution of the system of partial differential equations (\ref{eq:maxwell}) towards a state which fulfills equation~(\ref{eq:force_free_crossfield}) by introducing a suitable cross-field conductivity. In the numerical setup, we choose to combine the prescription of \citet{Komissarov2004} with the force-free current given above. This strategy effectively minimises the violations of equations\,(\ref{eq:force_free_crossfield}) and (\ref{eq:force_free_dominance}) by exponentially damping the (numerically induced) components of the electric field parallel to $\vect{\BB}$ and suitably adjusting the electric field in magnetospheric current sheets in order to obtain $\vect{\BB}^2-\vect{\EE}^2\rightarrow 0$ at these locations.

Throughout the literature, the magnetic dominance condition (\ref{eq:force_free_dominance}) condensates to a necessary condition of FFE \citep[e.g.][]{Uchida1997,McKinney2006}. For some authors \citep[e.g.][]{McKinney2006} the breakdown of the magnetic dominance implies the invalidity of the numerical model. Others \citep[e.g.][]{Uchida1997} claim that some physical processes (e.g. radiation losses) taking place in the regions where condition (\ref{eq:force_free_dominance}) is breached may restore the magnetic dominance condition. Indeed, \cite{Uchida1997} explicitly allows for transient phases violating condition (\ref{eq:force_free_dominance}) - these regions are then interpreted as abandoning the freezing of magnetic flux onto the flux of matter, being necessarily accompanied by dissipation. Following \citet{Uchida1997}, the force-free regime continues to be a valid approximation as long as the dissipative effects are only a small fraction of the total energy. The violation of the perpendicularity condition (\ref{eq:force_free_crossfield}) is an additional source of (Ohmic) dissipation \citep[studied for example in the context of Alfvén waves in force-free electrodynamics by][]{Li2018}. In practice, this channel of dissipation occurs when $\vect{\EE}\cdot\vect{\BB}\neq 0$ such that $\vect{\JJ}\cdot\vect{\EE}\neq 0$. Currently used force-free codes aim to avoid the transient into this regime by numerically cutting back all violations of condition (\ref{eq:force_free_dominance}) \citep[e.g.][]{Palenzuela2010,Paschalidis2013,Carrasco2016} or include a suitable Ohm's law \citep[e.g.][]{Komissarov2004,Spitkovsky2006,Alic2012,Parfrey2017} in order to minimise these violations during a transient phase. Figure~\ref{fig:energy_dissipation_channels} shows the breakdown of condition (\ref{eq:force_free_dominance}) during the simulation and hints towards the aforementioned dissipative processes. We refer to appendix \ref{sec:ff_breakdown} as well as, for example, \citet{Lyutikov2003} for further details on the necessary constraint preservation and limitations of the highly magnetised regime (such as the lack of physical reconnection). We will give a thorough review of the procedures employed in our code in a subsequent technical paper. 
	
\section{Twisted magnetar magnetosphere models}
\label{sec:twisted_models} 

\subsection{Magnetospheres}
\label{sec:initial_data} 

%
\begin{table}
	\centering
	\caption{Conversion table between code output in Heaviside-Lorentz \textit{geometrised} units ($M_\odot=G=c=1$) and \textit{non-geometrised} Gaussian units. In order to transform the respective quantities from code quantities to the \textit{non-geometrised} system, one has to multiply the \textit{geometrised} quantity by its conversion factor expressed in CGS.}
\label{tab:unit_conversions}
	{\renewcommand{\arraystretch}{1.4}
	\begin{tabular}{ccc}
		\hline
		Quantity & Nongeometrised unit & Conversion factor \\\hline
		Mass & M & $M_\odot$ \\ 
		Length & L & $M_\odot G c^{-2}$ \\ 
		Time & T & $M_\odot G c^{-3}$ \\
		Electric charge & $\text{L}^{3/2}\text{M}^{1/2}\text{T}^{-1}$ & $(4\pi)^{-1/2}M_\odot G^{1/2}$ \\ 
		Electric field & $\text{L}^{-1/2}\text{M}^{1/2}\text{T}^{-1}$ & $(4\pi)^{1/2}M_\odot^{-1} G^{-3/2}c^{4}$ \\ 
		Magnetic field & $\text{L}^{-1/2}\text{M}^{1/2}\text{T}^{-1}$ & $(4\pi)^{1/2}M_\odot^{-1} G^{-3/2}c^{4}$ \\ 
		Current density & $\text{L}^{-1/2}\text{M}^{1/2}\text{T}^{-2}$ & $(4\pi)^{-1/2}M_\odot^{-2} G^{-5/2}c^{7}$ \\ 
		(EM) Energy & $\text{L}^{2}\text{M}\:\text{T}^{-2}$ & $ M_\odot\:c^{2}$ \\
		(EM) Stress & $\text{L}^{-1}\text{M}\:\text{T}^{-2}$ & $ M_\odot^{-2} G^{-3}c^{8}$ \\ \hline
	\end{tabular}}
\end{table}
\begin{table}
 \centering
 {\renewcommand{\arraystretch}{1.5}
	\begin{tabular}{cccccccc}
		\hline
		& $s$ & $\sigma$ & $\Pc$ & $\ee/\ee_{\rm d}$ & $\JJ_{\rm max}$ & $\tilde{T}^{r\varphi}_{\rm max}$ & $\tilde{T}^{r\theta}_{\rm max}$ \\\hline
		A1 & 2 & 2 & 0.3294 & 1.1553 & 1.71e-6 & 8.97e-10 & 1.44e-9 \\
		A2 & 2 & 2 & 0.3303 & 1.3356 & 1.58e-6 & 8.95e-10 & 1.24e-9 \\ \hline
		B1 & 1 & 1 & 0.3717 & 1.1547 & 1.08e-6 & 7.68e-10 & 1.39e-9 \\
		B2 & 1 & 1 & 0.3720 & 1.2276 & 1.07e-6 & 7.68e-10 & 1.31e-9 \\\hline
		C1 & 1 & 1 & 0.4400 & 1.0653 & 1.95e-6 & 6.68e-10 & 1.56e-9 \\
		C2 & 1 & 1 & 0.4412 & 1.1943 & 1.03e-6 & 6.68e-10 & 1.44e-9 \\
		C3 & 1 & 1 & 0.4396 & 1.2738 & 1.03e-6 & 6.71e-10 & 1.35e-9 \\\hline
 \end{tabular}}
 \caption{Overview of initial data models used in our simulations. $s$, $\sigma$ and $\Pc$ are the parameters determining the boundary condition at the surface of the neutron star (see section~\ref{sec:initial_data}). $\ee$ denotes the total electromagnetic energy in the magnetospheres, which is normalised to the vacuum dipole energy $\ee_{\rm d}$ (equation\,\ref{eq:edipole}), hence without dimension. $\JJ_{\rm max}$ denotes the maximum current density at $t=0$ (see section~\ref{sec:current_magnetsophere} as well as Table~\ref{tab:unit_conversions} for unit conversion). The maximum initial electromagnetic stresses on the magnetar surface (equation\,\ref{eq:ffstress}) at $t=0$ are shown in the last two columns (i.e. $\TT^{ra}_{\rm max}:=\max_{\{|\vect{x}|=R_*\}}\{\TT^{ra}(t=0, \vect{x})\}$, with $a=\theta$, $\varphi$).
 Values of $\JJ_{\rm max}$ and $\TT^{ra}_{\rm max}$ are given in HLG units for a NS with $B_{\rm pole}=10^{15}$\,G and $R_*=\Rstar$\,km.}
 \label{tab:model_overview}
\end{table}
\begin{figure}
	\centering
	\includegraphics[width=0.45\textwidth]{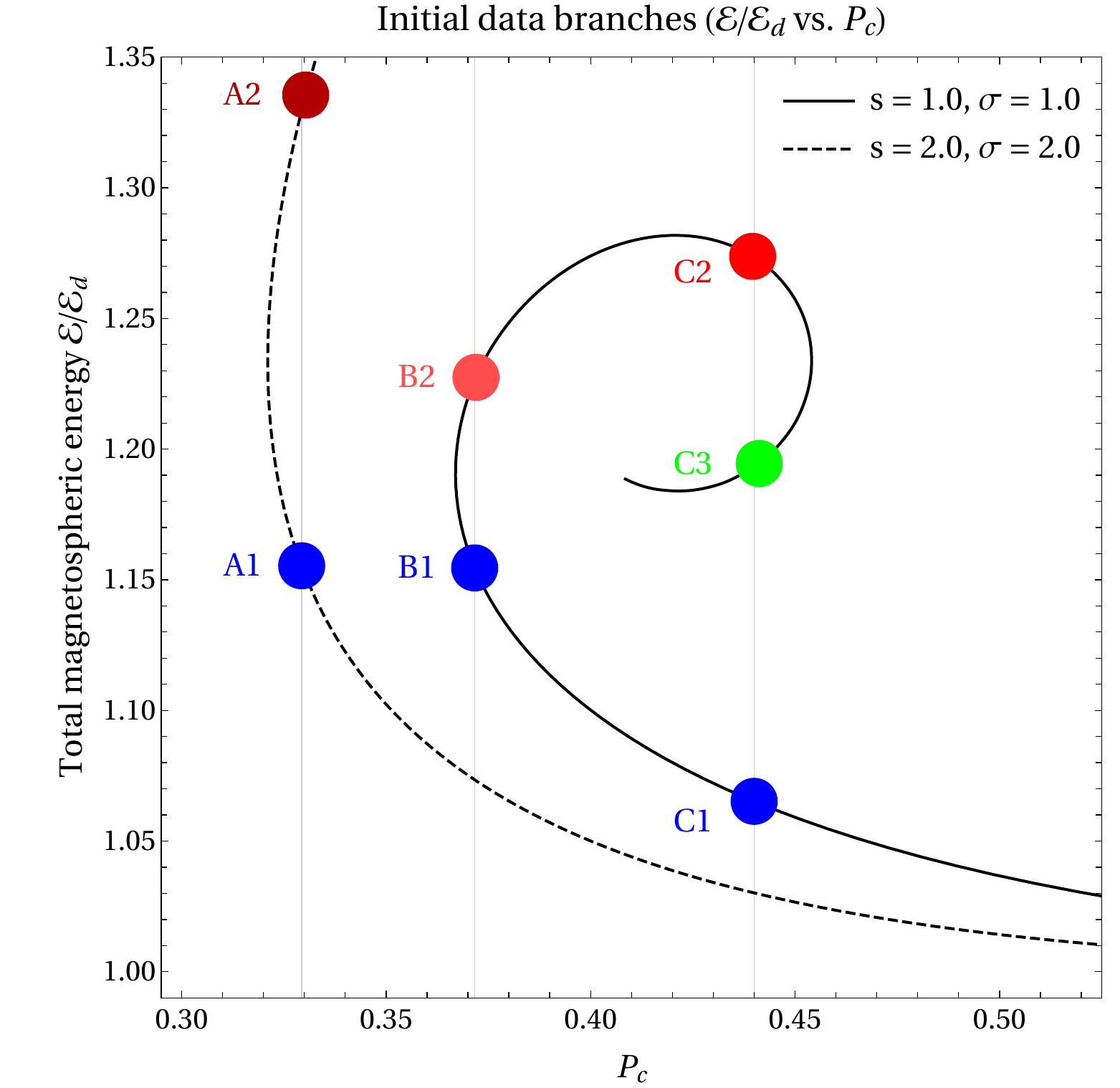} 
	\caption{Magnetospheric energy normalised to the vacuum dipole energy (equation~\ref{eq:edipole}) of the initial equilibrium models, for different values of the parameter $\Pc$ (in units of $P$ at the equator). The solid and dashed lines correspond to a series of models with constant $s$ and $\sigma$. The colored dots correspond to the initial data models used in our simulations.}
	\label{fig:initial_model}
\end{figure}

Due to the long rotational period of observed magnetars pushing the location of the light cylinder to great distances, it is possible to neglect the rotation of the neutron star when building numerical models of magnetospheres in the near zone. The equilibrium structure of a non-rotating axisymmetric force-free magnetosphere is given through the well-known GSE \citep{Luest1954,Grad1958,Shafranov1966}. This approach has been followed in several recent papers 
\citep[e.g.][]{Spitkovsky2006,Beskin2010,Vigano2011,Glampedakis2014,Fujisawa2014,Pili2015,Akgun2016,Akgun2018a,Kojima2017,Kojima2018,Kojima2018b}.
In most of these works, the toroidal field is confined within a magnetic surface near the equator, smoothly transitioning to vacuum at large distances. In stationary, non-rotating, axisymmetric magnetosphere models, the toroidal field cannot extend to the poles. Otherwise, the toroidal field would extend all the way to infinity, thus, violating the requirements of finite magnetic energy. Following the notation of \citet{Akgun2016,Akgun2018a}, we write the axisymmetric magnetic field in terms of its poloidal and toroidal components:
\begin{align}
 \vect{B} = \vect{\nabla} P \times \vect{\nabla}\varphi + T \vect{\nabla}\varphi \ ,
\end{align}
where $\varphi$ is the azimuthal angle in spherical coordinates. Here, $P$ and $T$ are the poloidal and toroidal stream functions.
Expressed in the orthonormal spherical basis corresponding to the coordinates $(r,\theta,\varphi)$, the magnetic field can be explicitly computed from the potentials $P$ and $T$ as
\begin{align}
B^r &= \frac{1}{r^2\sin\theta}\partial_\theta P, \label{eq:Br}\\ 
B^\theta &= - \frac{1}{r \sin\theta} \partial_r P, \label{eq:Btheta}\\
B^\varphi &= \frac{T}{r \sin\theta}. \label{eq:Bphi}
\end{align}
For an axially symmetric force-free field, the functions $T$ and $P$ may be expressed in terms of each other and appear as solutions of the force-free GSE:
\begin{align}
\left[\partial_r^2 + \frac{1 - \mu^2}{r^2}\partial_\mu^2\right] P + T \frac{dT}{dP} = 0,
\end{align}
where $\mu=\cos\theta$. $P$ and $T$ are constant on magnetic surfaces or, equivalently, along magnetic field lines. $P$ is related to the magnetic flux passing through the area centered on the axis and delineated by the magnetic surface. Therefore, its value at the poles is zero and increases towards the equator. The function $T$ is related to the current passing through the same area. Its functional dependence on $P$ can be chosen freely (consistently with any continuity and convergence requirements, particularly for the currents), which is equivalent to setting boundary conditions for $T$ at the surface of the star. Here, we invoke the same functional form for $T(P)$ as in \citet{Akgun2016,Akgun2018a}. Thus, the toroidal field is confined within some {\it critical} magnetic surface ($P = \Pc$),
\begin{align}
T\left(P\right)=\left\{\begin{array}{ccc}
s\times (P-\Pc)^\sigma& : & P \geqslant \Pc \\ 
0 & : & \text{else}
\end{array}\right.,
\label{eq:toroidal}
\end{align}
$s$ being a parameter determining the relative strength of the toroidal field with respect to the poloidal field. In order to avoid divergences in the currents we must demand that the power index satisfies $\sigma \geqslant 1$.
For a pure dipolar field, the poloidal stream function in the magnetosphere is
\begin{align}
P = \frac{1}{2}B_{\rm pole}\frac{R_*^3}{r}\sin^2\theta,
\end{align}
while the toroidal stream function is $T=0$ everywhere. We will consider the simplest cases where the boundary value of $P$ at the surface of the magnetar coincides with that of a dipolar field, and, therefore, the initial data is symmetric with respect to the equator. For different choices of the functional relation $T(P)$ given by equation~(\ref{eq:toroidal}) we solve the GSE and obtain a twisted magnetospheric initial model. We would like to note that all equations can be rescaled with $B_{\rm pole}$, hence, the results of our numerical simulations can be normalised to the field strength of interest.

The energy stored in the magnetosphere can be computed as a volume integral
 \begin{align}
 \ee = \frac{1}{8\pi}\int (\vect{B}^2 + \vect{E}^2)\, {\rm d}V .
 \end{align}
 For later reference and in order to normalize the energetic content of our models, we provide the energy stored in the magnetosphere of a pure dipolar magnetic field ($\vec{E}=0$, $B^r=B_{\rm pole}(R_*/r)^3\cos{\theta}$, $B^\theta=(B_{\rm pole}/2)(R_*/r)^3\sin{\theta}$, $B^\varphi=0$):
\begin{align}
\begin{split}
 \ee_{\rm d} = \frac{1}{12}B_{\rm pole}^2 R_*^3 =8.3\times 10^{46}\,\text{erg}\,\left(\frac{B_{\rm pole}}{10^{15}\,\text{G}}\right)^2\left(\frac{R_*}{10\,\text{km}}\right)^3.
\end{split}
 \label{eq:edipole}
 \end{align}
Once the surface value of $P$ and the functional relation $T(P)$ are defined, one can solve the GSE iteratively (as it is a non-linear equation), while imposing vacuum boundary conditions at large distances. We use the numerical code described in \citet{Akgun2018a} to build our initial data. Using this parametrisation, the boundary condition at the surface of the neutron star for the GSE (values of $P$ and $T$) is fully determined by four parameters $B_{\rm pole}$, $s$, $\Pc$ and $\sigma$. However, the solution of the GSE with this fixed boundary condition is not necessarily unique. \citet{Akgun2018a} showed that for sufficiently large magnetospheric twists, there exist degeneracies, i.e. different solutions of the GSE for the same boundary conditions (the same set of four parameters). These solutions differ in their energy, twist and the radial extent of the toroidal currents.

Table~\ref{tab:model_overview} shows the parameters used to construct the initial data for our numerical simulations. Each of the series A, B and C of initial models were chosen to have identical parameters but {\it different} magnetospheric energies and, hence, represent degenerate magnetospheric models. We would like to point out that the value of $\Pc$ is only equal, within each series, up to the second significant digit, due to numerical reasons. Figure~\ref{fig:initial_model} shows the energy of the initial models as a function of the parameter $\Pc$. Models within each spiral curve (constant $s$ and $\sigma$) and with the same value of $\Pc$ have identical boundary conditions but different energies. In the interpretation made by \citet{Akgun2018a}, the lower energy state for each series of degenerate models (i.e., A1, B1 and C1) corresponds to stable configurations, while high energy states (i.e., A2, B2, C2 and C3) may be unstable and would evolve towards the stable configuration releasing the respective energy difference. This instability is a possible scenario for the flare activity observed in magnetars.

The lowest energy solutions are the ones that are most similar to the vacuum solutions, with all field lines connected to the surface, while the higher energy solutions are more radially extended, and can contain disconnected field lines.

\begin{figure*}
	\centering
	\includegraphics[width=0.8\textwidth]{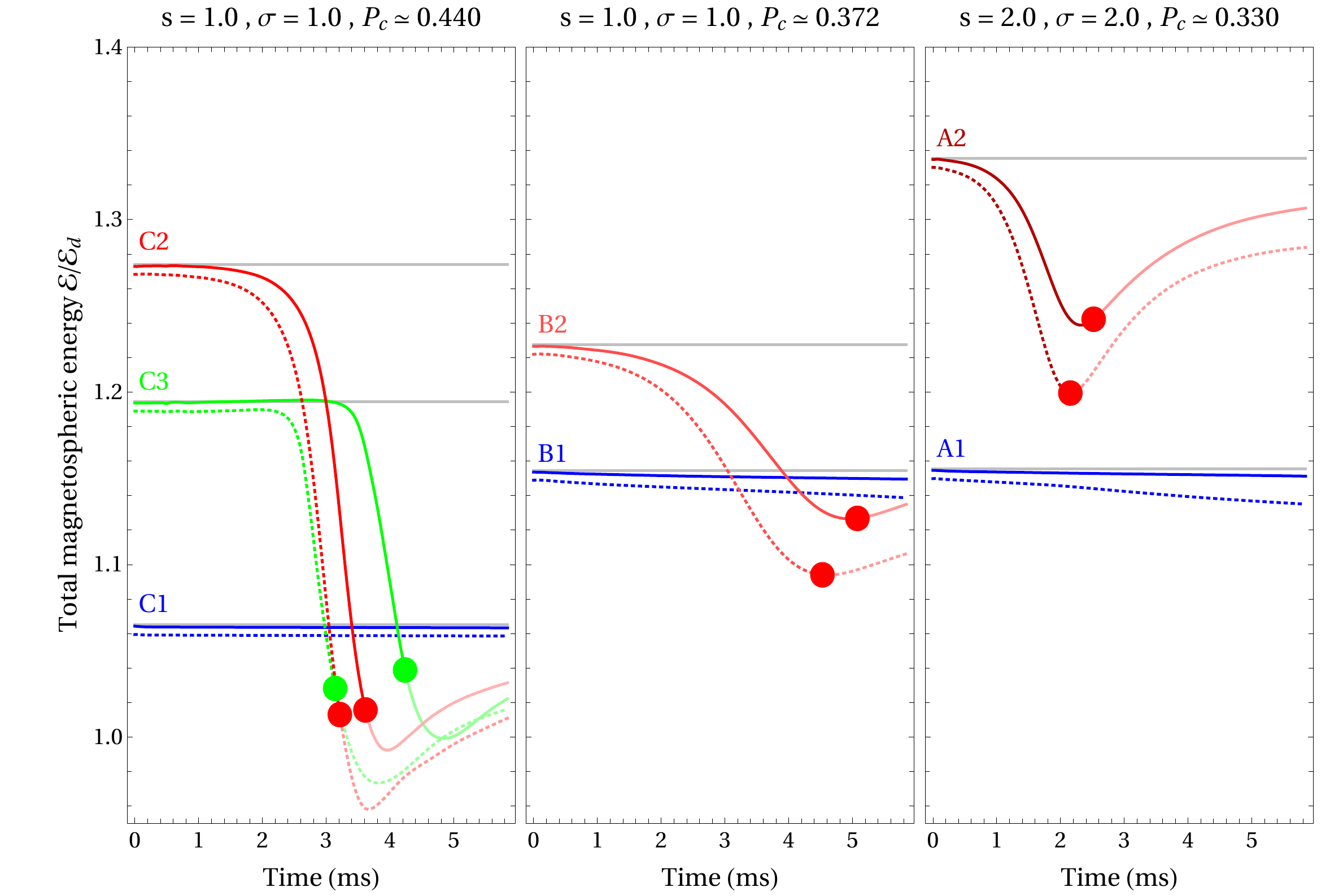} 
	\caption{Time evolution of total magnetospheric energy content for the models in Table~\ref{tab:model_overview}. The instability of the field configuration for degenerate solutions of higher energy triggers the rearrangement of magnetic field lines as well as a release of energy into the magnetosphere and onto the magnetar surface. The simulated timescale on which the instabilities are observed falls within the dynamical timescale of the magnetar crust. Low resolution simulations (16 points per $R_*$) are shown in dotted lines, high resolution simulations (32 points per $R_*$) in solid lines. The initial (analytical) value of total magnetospheric energy for each configuration is indicated by gray lines. The approximate time of the breakdown of the force-free condition $\vect{\EE}^2-\vect{\BB}^2<0$ (see appendix \ref{sec:ff_breakdown}) is depicted by colored dots.}
	\label{fig:absolute_energy_evolution}
\end{figure*}
%

\subsection{Magnetar interior}
\label{sec:interior} 

The initial models described above provide solutions only for the magnetosphere. For each possible magnetospheric model one can build infinite solutions to describe the neutron star interior. The magnetospheric (exterior) values of $P$ and $T$ determine the magnetic field $\vect{B}$ at the exterior side of the surface (equations~\ref{eq:Br} to~\ref{eq:Bphi}). To match this solution to the interior, one has to ensure the continuity of $B^r$ at the surface. This is valid if $P$ is continuous and, hence, $T$ and $B^\varphi$ are continuous as well. However, $B^\theta$ does not necessarily match continuously to the neutron star interior because current sheets (thin current-carrying layers across which the magnetic field changes either direction or magnitude) in the $\varphi$ direction may occur. Even if all components of $\vect{B}$ are continuous at the surface, the magnetic field structure in the interior depends completely on how currents are internally distributed.

In the astrophysical scenario we are considering, the magnetar reaches the initial state in which we start our numerical simulation after a slow magnetothermal evolution that proceeds in a long timescale compared to the dynamical timescales (cf. section~\ref{sec:timecales}) of the magnetosphere ($\sim 1$\,ms) or the crust ($\sim 10$\,ms). On such long timescales, any current close to the surface of the NS is expected to be dissipated by Ohmic diffusion. Therefore, we consider that initially all fields are continuous across the surface. We build our interior solution by extrapolating the exterior magnetic field towards the stellar interior across a number of grid cells as needed by the reconstruction algorithm used for the magnetospheric evolution in our simulations. Since the neutron star is basically a perfect conductor, the initial charge density and electric field in the interior (and the magnetosphere) are set to zero.

The surface values of $B^r$ and $B^\varphi$ are coincident for degenerate models (e.g. within the series C1, C2 and C3 in Figure~\ref{fig:initial_model}) because $P$ and $T$ at the surface are identical. However, since $P$ and $T$ may have a different radial dependence outside of the magnetar, and $B^\theta$ depends on the radial derivative of $P$ (equation~\ref{eq:Btheta}), it is different for every model of the same series.

\section{Simulations}
\label{sec:simulations} 

We have performed numerical simulations of the neutron star magnetosphere using the initial models in Table~\ref{tab:model_overview}. For all the simulations we employ our own implementation of a General Relativistic FFE code in the framework of the \texttt{Einstein Toolkit}\footnote{\url{http://www.einsteintoolkit.org}} \citep{Loeffler2012}. The \texttt{Einstein Toolkit} is an open-source software package utilizing the modularity of the \texttt{Cactus}\footnote{\url{http://www.cactuscode.org}} code \citep{Goodale2002a} which enables the user to specify so-called \texttt{thorns} in order to set up customary simulations. There exist other code packages such as \texttt{GiRaFFE} \citep{Etienne2017}, which integrate the equations of force-free electrodynamics employing an evolution scheme based on the Poynting flux as a conserved quantity \cite[cf.][]{McKinney2006,Paschalidis2013} rather than the electric field and its current sources \cite[as formulated in, e.g.][]{Komissarov2004,Parfrey2017}. The \texttt{Einstein Toolkit} employs units where $M_\odot=G=c=1$, which sets the respective time and length scales to be $1M_\odot\equiv 4.93\times 10^{-6}\text{ s}\equiv 1477.98\text{ m}$. This unit system is a variation of the so-called system of \textit{geometrised units} \citep[as introduced in appendix F of][]{Wald2010}, with the additional normalisation of the mass to $1M_\odot$ (i.e. our HLG units, as introduced in section~\ref{sec:ff_equations}). For easy reference, we provide a set of conversion factors for relevant physical quantities in Table~\ref{tab:unit_conversions}.
\begin{figure*}
	\centering
	
	\includemedia[width=1.0\linewidth,
	height=0.791594\linewidth,
	activate=onclick,
	passcontext,
	transparent,
	addresource=images/figure4a.mp4,
	flashvars={flv=images/figure4a.mp4}
	]{\includegraphics[width=1.0\textwidth]{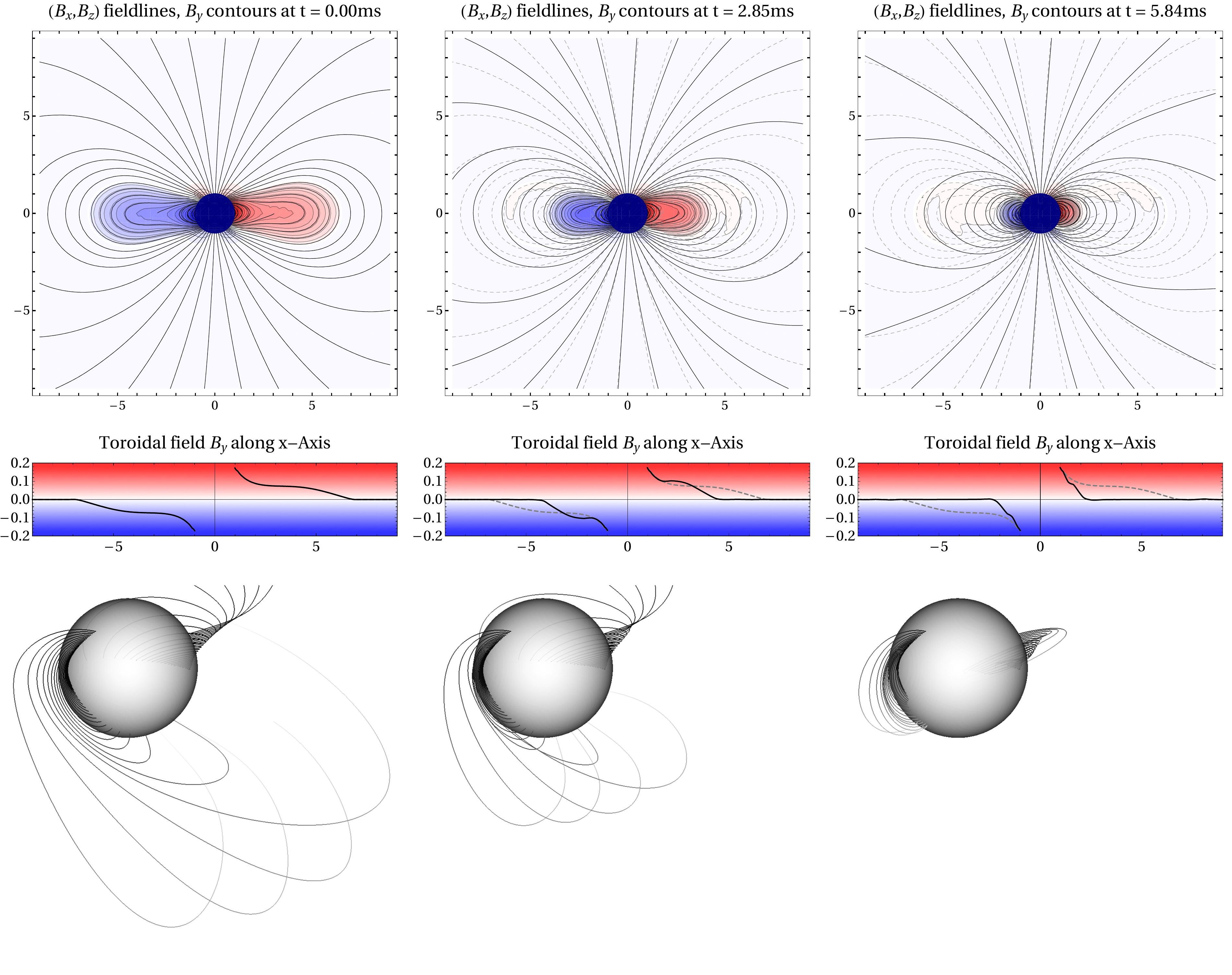}}{player.swf}
	\vspace{-11pt}
	\caption{Field line evolution (high resolution, 32 points per $R_*$) of the C2 initial model (Table~\ref{tab:model_overview}). The initially extended lobes of magnetic twist relax towards a dipolar structure and fall towards the central object. Strong energy dissipation (see Figure~\ref{fig:absolute_energy_evolution}) occurs when the magnetic twist collapses onto the magnetar crust. The final configuration is dipole-like, though it fully relaxes on a much longer dynamical timescale. \textit{Top:} Poloidal field lines (cross-section through the 3D data) and color-filled contours of the toroidal magnetic field (same color coding as below). The initial field line configuration is indicated by gray dashed lines. \textit{Middle:} Toroidal field distribution along the x axis. The initial toroidal magnetic field is denoted by gray dashed lines. \textit{Bottom:} Evolution of selected field lines in 3D, displaying the twist relaxation. \textit{Click for animation:} Evolution of total magnetospheric energy and selected field lines in 3D of the (high resolution) C2 initial model (Acrobat Reader only).}
	\label{fig:high_energy_relaxation}
\end{figure*}
%
\subsection{Numerical setup}
\label{sec:numerical_setup} 

All shown simulations are conducted on a 3D box with dimensions $\left[4741.12 M_\odot\times4741.12 M_\odot\times4741.12 M_\odot\right]$ with a grid spacing of $\Delta_{x,y,z}=74.08M_\odot$ on the coarsest grid level. For the chosen magnetar model of radius $R_*=9.26M_\odot$ ($\simeq \Rstar$\,km) this corresponds to a $\left[512R_*\times 512R_*\times 512R_*\right]$ box with a grid spacing of $\Delta_{x,y,z}=8R_*$. For the low and high resolution tests we employ seven and eight additional levels of mesh refinement, each increasing the resolution by a factor of two and encompassing the central object, respectively. This means that the finest resolution of our models (close to the magnetar surface) are $\Delta_{x,y,z}^{\rm min}=0.0625\times R_* = 0.5787M_\odot$ and $\Delta_{x,y,z}^{\rm min}=0.03125\times R_*= 0.2894M_\odot$ for the low and high resolution models, or in other words 16 and 32 points per $R_*$, respectively. The initial data is evolved for a period of $t=1185.28M_\odot\simeq 5.84\,$ms, which is chosen to be well below the dynamical timescale of the magnetar crust, which can be considered as a fixed boundary (see section~\ref{sec:timecales}).

In order to ensure the conservation properties of the algorithm, it is critical to employ {\it refluxing} techniques correcting numerical fluxes across different levels of mesh refinement \citep[see, e.g.][]{Collins2010}. Specifically, we make use of the thorn \texttt{Refluxing}\footnote{Refluxing at mesh refinement interfaces by Erik Schnetter: \url{https://svn.cct.lsu.edu/repos/numrel/LSUThorns/Refluxing/trunk}} in combination with a cell-centered refinement structure \citep[cf.][]{Shibata2015}. We highlight the fact that employing the refluxing algorithm makes the numerical code $2-4$ times slower for the benefit of enforcing the conservation properties of the numerical method (specially of the charge). Refluxing also reduces the numerical instabilities, which tend to develop at mesh refinement boundaries.

In conservative schemes, numerical reconstruction algorithms \citep[we employ an MP7 scheme, cf.][]{Suresh1997} derive inter-cell approximations of the conservative variables by making use of their values at several adjacent grid-points (for MP7, one requires seven points). As a result of the numerical coupling between the magnetosphere and the magnetar crust introduced by the inter-cell reconstruction at the stellar surface, the field dynamics induce a mismatch in the current flowing through the surface and effectively trigger a (numerical) flow of charges leaving or entering the domain. In order to avoid this artifact, we replace the reconstructed values of the radial current $\JJ^r_{\rm FFE}$ at interfaces between the stellar interior and exterior by the cell-centered value in the stellar interior. This procedure ensures a conservation of magnetospheric charge.

The (3D) initial data is imported from the (2D) initial models (see se. \ref{sec:initial_data}) by bicubic spline interpolation. 
Throughout the numerical evolution, all quantities on grid-points inside of the magnetar radius are fixed to their initial values.

\subsection{Instability onset and magnetospheric energy balance}
\label{sec:instability_relax} 

%
\begin{figure}
	\centering
	\includegraphics[width=0.45\textwidth]{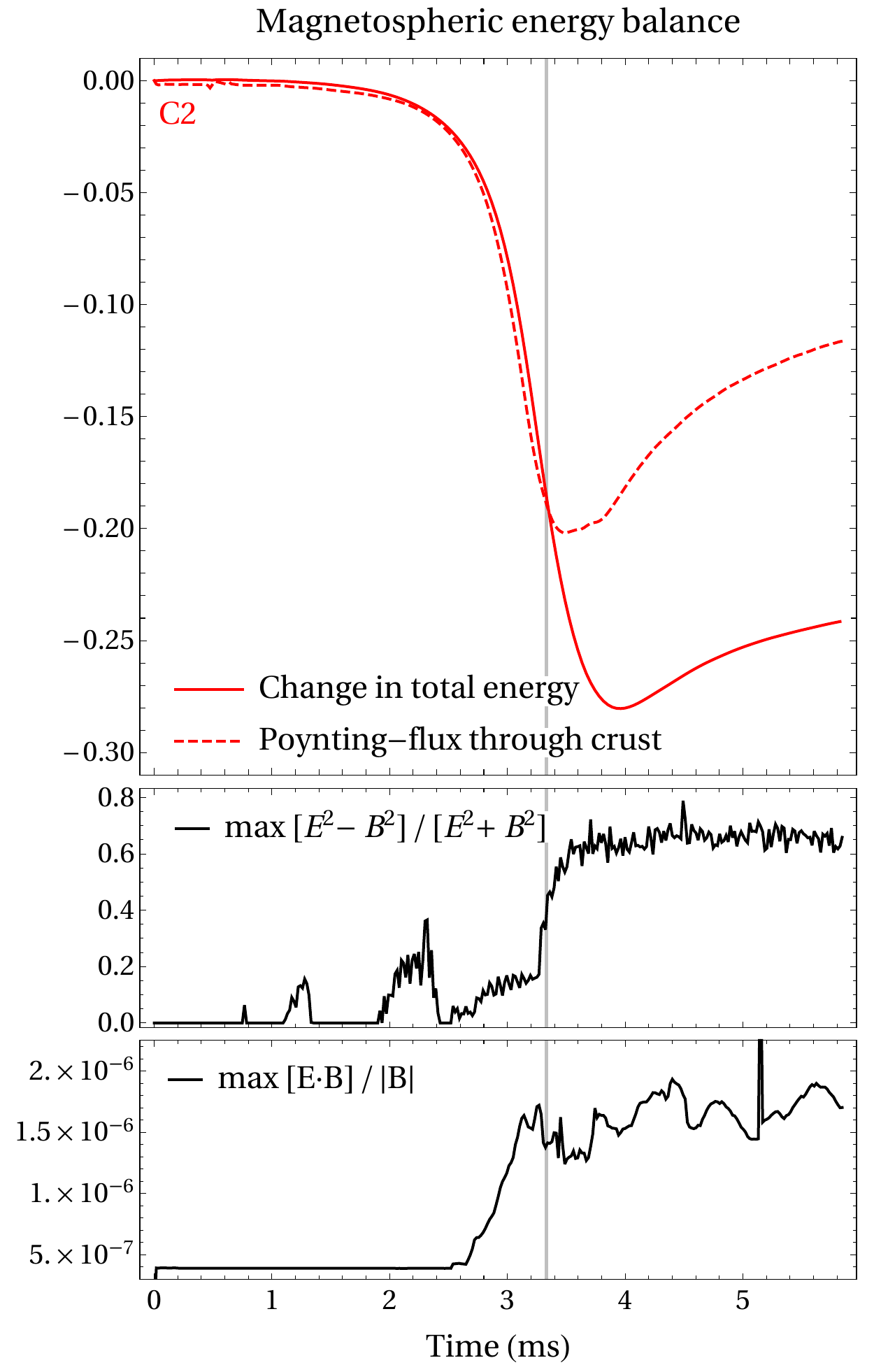} 
	\vspace{-6pt}
	\caption{Energy balance during the evolution of the high resolution model C2 (Table~\ref{tab:model_overview}). \textit{Top:} Comparison of the change in total magnetospheric energy, normalized to the energy of a magnetosphere equiped with a pure dipolar magnetic field, $\Delta \ee/\ee_{\rm d}$, as well as the Poynting flux through the magnetar surface. Up to a simulation time of $t~\sim 3.33\,$ms the energy change is dominated by Poynting flux onto the magnetar crust. \textit{Middle:} Maximum violation of the $\vect{\BB}^2-\vect{\EE}^2\ge 0$ condition throughout the numerical grid. \textit{Bottom:} Maximum violation of the $\vect{\BB}\cdot\vect{\EE}= 0$ constraint throughout the numerical grid. At the time of the breakdown of conditions (\ref{eq:force_free_crossfield}) and (\ref{eq:force_free_dominance}), the energy change is dominated by secondary (possibly numerical) effects. }
	\label{fig:energy_dissipation_channels}
\end{figure}

We have performed simulations with initial models in the low energy branch (A1, B1 and C1) and in the high energy branch (A2, B2, C2, C3). We observe a differentiated behavior in the evolution of the system depending on the class of initial model. For models in the low energy branch we find that the magnetosphere is stable and that the system remains essentially unchanged. The energy of the system remains constant throughout the simulation (see blue lines in Figure~\ref{fig:absolute_energy_evolution}), confirming the stability of these configurations, at least on dynamical timescales. This is specially true in the high resolution models, which exhibit a smaller numerical dissipation. The slightly larger numerical dissipation of the low resolution models explains the small drift in time with respect to the initial energy displayed by the blue dashed lines in Figure~\ref{fig:absolute_energy_evolution}. On the other hand, models in the high energy branch become unstable on a timescale of a few milliseconds and the magnetosphere changes its shape roughly at the same time as the energy of the magnetosphere decreases (see red and green lines in Figure~\ref{fig:absolute_energy_evolution}). This numerical experiment confirms the hypothesis of \citet{Akgun2018a} that, for degenerate initial models, only the lowest energy state is stable, and that all corresponding degenerate cases of high energy are unstable. In addition, we note that the lower energy states are closer to a purely dipolar magnetosphere, hence, the minimised circumference of the magnetic surfaces minimise the magnetospheric energy content \citep[cf.][]{Thompson1996}.

For configurations in the unstable branch, the onset of the instability proceeds earlier for lower numerical resolution. This is expected because a coarser grid contains larger numerical discretisation errors acting as a seed for the instability onset. However, the rapid drop in energy during the instability proceeds in a similar fashion for both numerical resolutions, indicating that the instability has a physical origin and is not a numerical artifact. In the case of the high energy initial model C2 we observe a rearrangement of the lobes of magnetic twist towards a dipolar structure (see Figure~\ref{fig:high_energy_relaxation}) prior to a significant drop of magnetospheric energy (by approximately $30\%$ of its initial value). During the phase of full validity of the force-free condition (see equation~\ref{eq:force_free_dominance}) the loss of magnetospheric energy is dominated by an outgoing Poynting flux at the innermost boundary (see Figure~\ref{fig:energy_dissipation_channels}). For our boundary condition it can be interpreted as the formation of a strong current on a thin layer below the surface, where energy can be efficiently dissipated.

Following \citet{Parfrey2013} in the context of twisted magnetar fields and \citet{Li2018} in a study of energy dissipation in collisions of force-free Alfvén waves, the onset of the (topological) relaxation is likely to be linked to Ohmic heating $\vect{J}\cdot\vect{E}\neq 0$, which occurs as a result of (minor) violations of the force-free condition (\ref{eq:force_free_crossfield}), as can be seen in the bottom panel of Fig.~\ref{fig:energy_dissipation_channels} (note the much smaller scale of that panel compared to the middle one). We give a more detailed review of the treatment of these violations in our code and throughout the literature in appendix \ref{sec:ff_breakdown}.

\subsection{Surface currents and long-time evolution}
\label{sec:surface_currents} 

\begin{figure}
	\centering
	\includegraphics[width=0.49\textwidth]{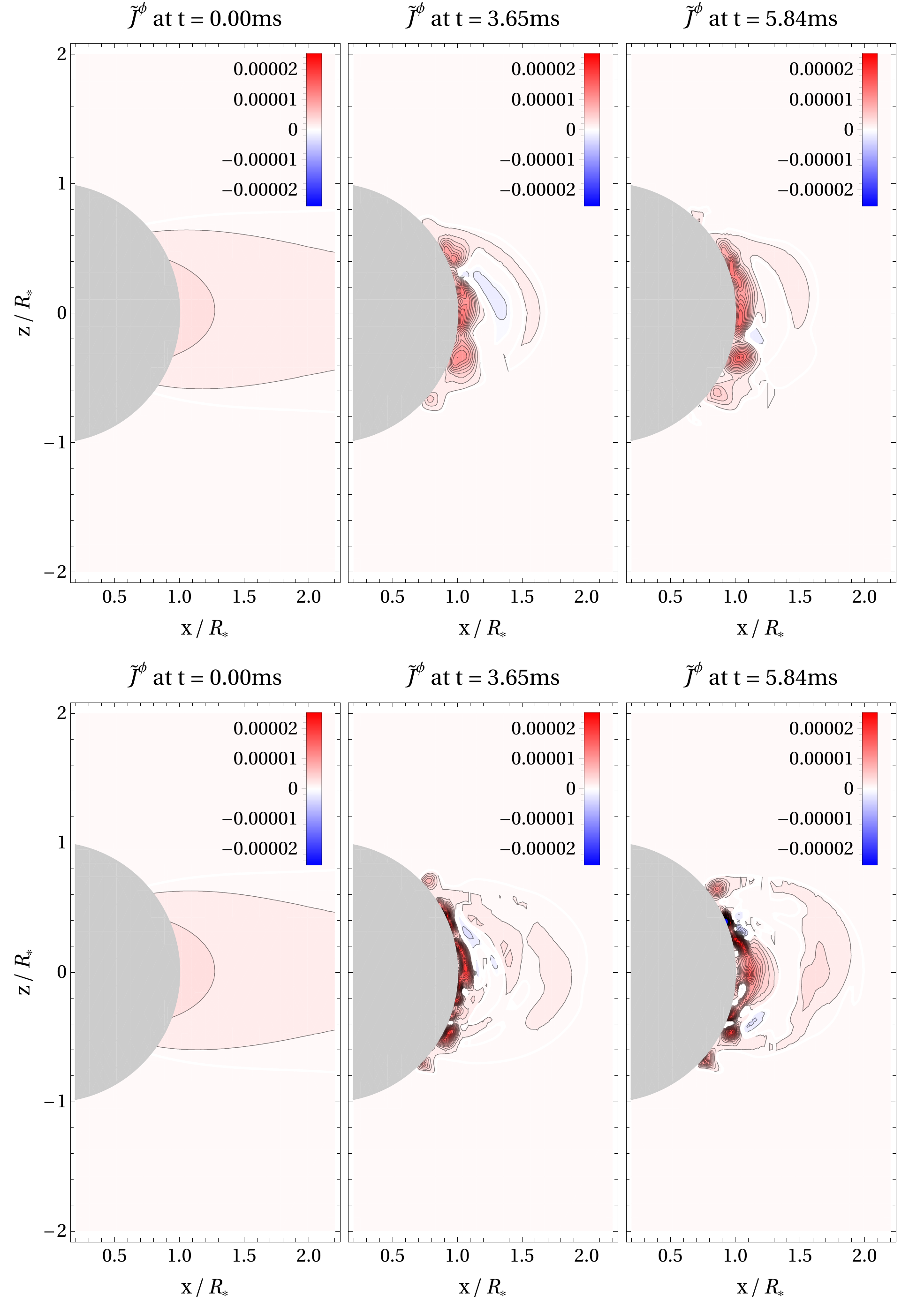} 
	\vspace{-11pt}
	\caption{\textit{xz}-cross-sections of the toroidal current in geometrised units showing the development of strong surface currents during the evolution, in addition to other currents extended on larger magnetospheric volumes. \textit{Top:} Low resolution model C2 (16 points per $R_*$). \textit{Bottom:} High resolution model C2 (32 points per $R_*$). The high resolution evolution shows currents located around the magnetar surface with more detailed structures, emphasizing their interpretation as surface currents. The spatial coincidence of the currents in both resolutions reinforce the argument that the observed currents are of physical origin (in spite of the - relatively small - differences among different resolutions).}
	\label{fig:surface_currents_xz}
\end{figure}

\begin{figure}
	\centering
	\includegraphics[width=0.45\textwidth]{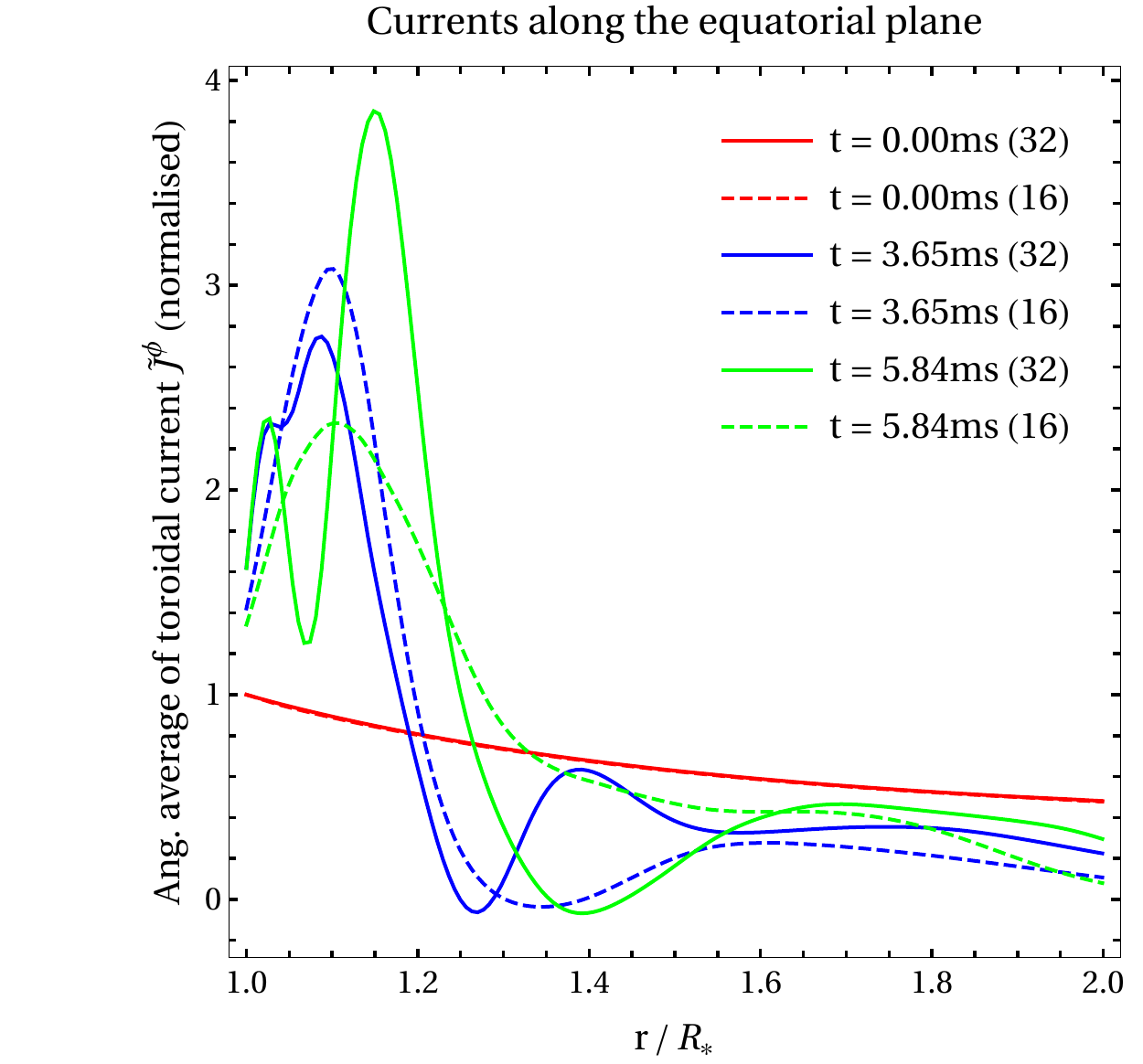} 
	\vspace{-6pt}
	\caption{Azimuthal angular averages of the toroidal current (normalised to its initial value at the stellar surface) in the equatorial plane showing the development of surface currents during the evolution of the C2 initial model. We display the current evolution for both low resolution (16 points per $R_*$, denoted by dashed lines), and high resolution (32 points per $R_*$, denoted by solid lines) models. The increase of the toroidal current during the transient of energy dissipation (see Figure~\ref{fig:absolute_energy_evolution}) at the lower resolution (compare the two blue lines) may be attributed to a faster onset of the twist instability for this model.}
	\label{fig:surface_currents_avg}
\end{figure}

Following the initial instability and subsequent rapid rearrangement of the magnetar magnetosphere (section~\ref{sec:instability_relax}), thin currents form at the magnetar surface (see Figures~\ref{fig:surface_currents_xz} and \ref{fig:surface_currents_avg}). These currents are expected to appear as the initial model in the high energy state tries to relax to the lowest energy magnetospheric configuration, while keeping the interior field fixed (see the discussion in section~\ref{sec:interior}). There are two possible fates for these currents: i) they could propagate inwards, inside the magnetar crust, deforming the magnetic field inside and creating a mechanical stress in the crust, on a timescale of several $10 \text{ ms}$, or ii) they could form a thin surface current dissipating on a timescale shorter that the time it takes to deform the crust. These two possibilities are not mutually exclusive and a combination of both is possible. In none of the cases our simulations can give a conclusive answer because i) we are not evolving the magnetar interior as we are considering only timescales smaller than the dynamical timescale of the crust, ii) the formation of thin surface currents is numerically challenging (would require a computationally prohibitively high resolution near the magnetar surface), and iii) it would eventually violate the FF conditions (\ref{eq:force_free_crossfield}) and (\ref{eq:force_free_dominance}), hence invalidating our current numerical approach.

The aforementioned current layers are expected to be regions of strong energy dissipation and the breakdown of the force-free conditions \citep[see, e.g.][]{Uchida1997,McKinney2006,Palenzuela2010,Parfrey2013}. Figures~\ref{fig:energy_dissipation_channels} and~\ref{fig:surface_currents_avg} link the breakdown of the force-free condition (\ref{eq:force_free_dominance}) and the occurrence of surface currents with the opening of dissipation channels different to the energy flow through the magnetar surface (see appendix \ref{sec:ff_breakdown} for a short review of the force-free breakdown). We find the violation of condition (\ref{eq:force_free_crossfield}) to be continuously occurring with peaks
at the instance of rapid energy dissipation. Condition (\ref{eq:force_free_dominance}) starts to fail on longer timescales at the moment of fastest transfer of magnetic energy through the surface. At this time, further dissipation mechanisms (see Figure~\ref{fig:energy_dissipation_channels}) come into play, as is expected throughout the literature \citep{Uchida1997,McKinney2006,Li2018}.

It should be noted that the total magnetospheric energy for the models B2, C2, and C3 drops below the energy of their respective low energy equilibrium solutions, and even below the magnetospheric energy of the vacuum dipole (equation~\ref{eq:edipole}). However, this energy drop is (slightly) smaller for the high resolution simulations, and shows some dependence on the chosen setup of the hyperbolic/parabolic cleaning procedures (see appendix \ref{sec:augmented_system}) at the magnetar surface. The sensitivity of this behavior to the numerical details at the location of the (3D Cartesian) crust may be attributed to the numerical dissipation of the employed code.

\begin{table}
	\centering
 \caption{Selection of electromagnetic quantities monitored throughout the (high resolution, 32 points per $R_*$) simulation of the models of Table~\ref{tab:model_overview}. The total change in energy $\Delta \ee$ (displayed as a fraction of the vacuum dipole energy; equation~\ref{eq:edipole}) corresponds to the maximum drop of electromagnetic energy during the total runtime (see section~\ref{sec:energy_release}). The operator $\Delta_{\rm mx}$ acting on any quantity $A(t,\vect{x})$ is defined as $\Delta_{\rm mx}A:=\max_{\{t,|\vect{x}|=R_*\}}{ \{ A(t,\vect{x})-A(0,\vect{x}) \}/\max_{\{|\vect{x}|=R_*\}}A(0,\vect{x}) }$. Hence, $\Delta_{\rm mx} J$ is the maximum increase in current density in the magnetosphere during the relaxation relative to the initial values (see section~\ref{sec:current_magnetsophere}). In the right columns, $\Delta_{\rm mx} T^{r\varphi}$ and $\Delta_{\rm mx} T^{r\theta}$ denote the maximum increase of electromagnetic stresses relative to their corresponding initial values (see section~\ref{sec:crust_stresses}) on the stellar surface compared to its initial value. We highlight with bold face the maximum values of each of the last four columns.}
\label{tab:model_results}
	{\renewcommand{\arraystretch}{1.5}
		\begin{tabular}{cccccc}
			\hline
			Model & $\Delta t_{\rm r} \text{(ms)}$ & $\Delta \ee/\ee_{\rm d}$ & $\Delta_{\rm mx} J$ & $\Delta_{\rm mx} T^{r\varphi}$ & $\Delta_{\rm mx} T^{r\theta}$ \\*[0pt]\hline
			A1 & 5.8400 & 0.0033 & 0.0159 & 0.0012 & 0.0010 \\
			A2 & 1.4162 & 0.0963 & 1.6350 & 0.0295 & 0.0150 \\\hline
			B1 & 5.8400 & 0.0042 & 0.0363 & 0.0012 & 0.0014 \\
			B2 & 3.0427 & 0.1002 & 0.9805 & 0.0358 & 0.0232 \\\hline
			C1 & 5.8400 & 0.0009 & 0.0640 & 0.0008 & 0.0013 \\
			C2 & 2.1604 & \textbf{0.2808} & \textbf{3.5400} & 0.0851 & 0.0414 \\
			C3 & 1.0490 & 0.1962 & 3.1720 & \textbf{0.1008} & \textbf{0.0811} \\\hline
		\end{tabular}}
\end{table}
%

%
\section{Discussion}
\label{sec:discussion} 


\subsection{Energy release during the instability}
\label{sec:energy_release} 
During the rearrangement of magnetic field lines in the high energy
models A2, B2, C2, and C3, an amount $\Delta \ee$ of electromagnetic
energy is released into the magnetosphere and onto the magnetar crust
(Poynting flux through the stellar surface, see
Figure~\ref{fig:energy_dissipation_channels}). The amount of released
energy in CGS units, $\ee_r$, can be quantified directly from
Table~\ref{tab:model_results} by employing the conversion formula
\begin{align}
\ee_{\rm r}=\:2.14\times 10^{47}\,\text{erg}\,\left(\frac{\Delta \ee}{\ee_{\rm d}}\right)\left(\frac{B_{\rm pole}}{10^{15}\text{
 G}}\right)^2 \left( \frac{R_*}{\Rstar}\right)^3\,.
\end{align}
For the changes in energy
($\Delta \ee/\ee_{\rm d} \approx 0.1 - 0.3$) observed in our
simulations with the highest energy within each series (C2,
 C3, B2 and A2) the released energy is in the range
$\ee_{\rm r}\approx 2.1\times 10^{46}-6.4\times
 10^{46}$\,erg.
 This energy range is compatible with that of observed GFs
 ($10^{45}-10^{48}$\,erg). For instance, the energy liberated during
 the peak of the GF of SGR\,1806-20 is $\sim 3.7\times 10^{46}\,$erg
 \citep{Hurley2005}, which is compatible with values
 $\Delta \ee/\ee_{\rm d}\simeq 0.17$. However, the other two known GF
 events \citep[SGR\,0525-66 and SGR\,1900+14, see][]{Cline1980,Hurley1999} display significantly smaller amounts of energy during their
 initial peaks.

The range of $\Delta \ee/\ee_{\rm d}$ in our
simulations depends on the choice of initial models. The detailed
analysis in \citet{Akgun2018a} shows that $\Delta \ee/\ee_{\rm d}$ could
in principle be as large as $0.8$ for models with the appropriate
values of $s$ and $\sigma$ and the value of $\Pc$ to be at the maximum
of the corresponding sequence \citep[see Figure~3
in][]{Akgun2018a}. However, the astrophysical path that could lead to
an unstable configuration this far away from the equilibrium branch is
unclear. Speaking in terms of evolution, models close to the stability
threshold for which $\Delta \ee/\ee_{\rm d}$ could be a small
fraction of the energy encountered in our simulations are much more
likely than models with values of, e.g. $\Delta \ee
/\ee_{\rm d}> 0.2$.

The timescale on which $\ee_{\rm r}$ is released
($\Delta t_{\rm r}\sim 1-5$\,ms; see Table~\ref{tab:model_results}) is consistent with the dynamical
 timescales in the magnetosphere (section~\ref{sec:dynamical}). If we
 estimate the luminosity of the energy released as
 \begin{align}
 L_0 := \frac{\ee_{\rm r}}{\Delta t_{\rm r}}\: ,
 \label{eq:L0}
\end{align}
we find that $L_0 \sim (0.7 - 4) \times 10^{49}\,$erg\,s$^{-1}$ for
the unstable models listed in
Table~\ref{tab:effective_temperature}. This dynamical luminosity is
significantly larger than the peak luminosity of GFs (e.g. the peak
luminosity of SGR\,1806-20 is $\sim 2\times 10^{47}\,$erg\,s$^{-1}$;
\citealt{Hurley2005}), and suggests that only a fraction of
 the released energy contributes to explain the thermal properties
 of GFs in SGRs. As an alternative, not necessarily
 exclusive, we consider different mechanisms to broaden the time
 scale over which the energy leaks out of the system, hence reducing
 $L_0$, in the following sections.

\subsection{Stresses induced in the crust}
\label{sec:crust_stresses} 

Figure~\ref{fig:energy_dissipation_channels} suggests that a
significant part of the released energy is transferred into the
magnetar crust during the (fully force-free) evolution. We would like
to point out that an exact modeling of magnetar crust physics will be
necessary in order to simulate respective feedback mechanisms between
the stellar surface and the magnetosphere. However, in this section
 we make some crude estimates regarding the stresses induced in the
 crust as a result of the magnetospheric evolution of our models.

The stresses induced in the crust by the evolving magnetosphere can be computed
studying the momentum-transfer from the magnetosphere to the crust. The stress tensor 
in the (force-free) magnetosphere is
\begin{align}
T_{\rm ms}^{ij} = \frac{1}{4\pi}\left(\frac{1}{2} \delta^{ij}\left(E_{\rm ms}^2+B_{\rm ms}^2\right)- E_{\rm ms}^i E_{\rm ms}^j- B_{\rm ms}^i B_{\rm ms}^j\right),
 \label{eq:ffstress}
\end{align}
where $B^i_{\rm ms}$, and $E^i_{\rm ms}$ are the magnetic and electric fields in the magnetosphere.
The stress tensor in the crust consists of the contribution of the magnetic field, the fluid,
and the stress of the solid
\begin{align}
T_{\rm c}^{ij} = \mathcal{P} \delta^{ij} + \frac{1}{4\pi} \left( \frac{1}{2}
 \delta^{ij} B_{\rm c}^2- B_{\rm c}^i B_{\rm c}^j\right)+ \sigma^{ij},
\end{align}
where $\mathcal{P}$ is the pressure of the fluid, $B^i_{\rm c}$ the magnetic field inside the crust and 
$\sigma^{ij}$ is the stress tensor of the deformed solid. Especially, $\sigma^{ij}=0$ for a non-deformed solid - which holds at the beginning of the presented simulations in which the crust is relaxed after the long-term magneto-thermal evolution during which plastic deformations can keep this relaxed state. Throughout the instability phase captured in our simulations, the magnetosphere induces
a stress in the crust that effectively deforms it. The Lagrangian displacement of any point in the crust with respect to the relaxed state is given by the deformation vector $\xi^i$. For linear displacements, the stress tensor can be expressed
in terms of the deformation vector \citep{Landau:elasticity} as follows:
\begin{align}
\sigma^{ij} = K \xi^k_{\,;k} f^{ij}+ 2\mu \left (\frac{1}{2} (\xi^{j\,;i} + \xi^{i\,;j}) - \frac{1}{3} f^{ij} \xi^k_{\,;k}\right ),
\end{align}
where semicolon indicates the covariant derivative, $f_{ij}$ the flat 3-metric, 
$K$ is the bulk modulus and $\mu$ the shear
modulus. Crust and magnetosphere can only interchange momentum through
$T^{r\theta}$ and $T^{r\varphi}$. Hence, these are the only relevant components. Imposing continuity of
these two components at the surface of the star ($\mathcal{P}=0$) one finds
\begin{align}
- \frac{1}{4\pi}\left( E_{\rm ms}^r E_{\rm ms}^a+ B_{\rm ms}^r B_{\rm ms}^a\right)=-\frac{1}{4\pi} B_c^r B_c^a + \sigma^{ra}\qquad a=\{\theta,\varphi\},
\end{align}
and therefore
\begin{align}
\sigma^{ra} =\frac{1}{4\pi}\left( B^r_{\rm c} B^a_{\rm c} - E_{\rm ms}^r E_{\rm ms}^a - B_{\rm ms}^r B_{\rm ms}^a\right) \qquad a=\{\theta,\varphi\}. \label{eq:stress}
\end{align}
For the equilibrium configuration at the beginning
of the simulation, in which $\vect{E}=0$ and $\vect{B}$ is continuous (no initial current sheets),
the mechanical stress is zero ($\sigma^{ra}=0$) and, hence, the stress at the surface is just $T_{\rm c}^{ra}= - B_{\rm ms}^r(t=0) B_{\rm ms}^a (t=0)/(4\pi)$. Therefore, we can compute the mechanical stress at any time as
 \begin{align}
 \sigma^{ra} = T^{ra}_{\rm ms} - T^{ra}_{\rm ms} (t=0).
 \end{align}
As discussed in section~\ref{sec:dynamical}, the magnetic fields are dominant in the outermost low-density part of the crust and can be considered to be force-free \citep{Beloborodov2009}. The
point at which the magnetic field lines are anchored is not the
surface of the star, but some radius, $r_{\rm c}$, below it (see also the discussion referencing Figure~\ref{fig:speeds}). However, equation~(\ref{eq:stress}) still holds at this radius, because $P$ is continuous, and the relevant terms cancel out.
In other words, from the point of view of the numerical simulation,
the inner boundary condition therein used
corresponds to $r_{\rm c}$, and not the radius of the star. The force-free region of the crust
corresponds to the region where shear stresses do not play a role in the dynamics, i.e. $\mu \ll B^2$.
For typical magnetar magnetic fields of $B\sim 10^{15}$\,G this is
fulfilled for $\mu_{\rm c} \ll 10^{30}$\,erg\,cm$^{-3}$,
which typically and for a large variety of equations of state
\citep{Steiner2009} corresponds to densities of $\rho \ll 10^{14}$\,g\,cm$^{-3}$. 

\begin{figure}
	\centering
	\includegraphics[width=0.45\textwidth]{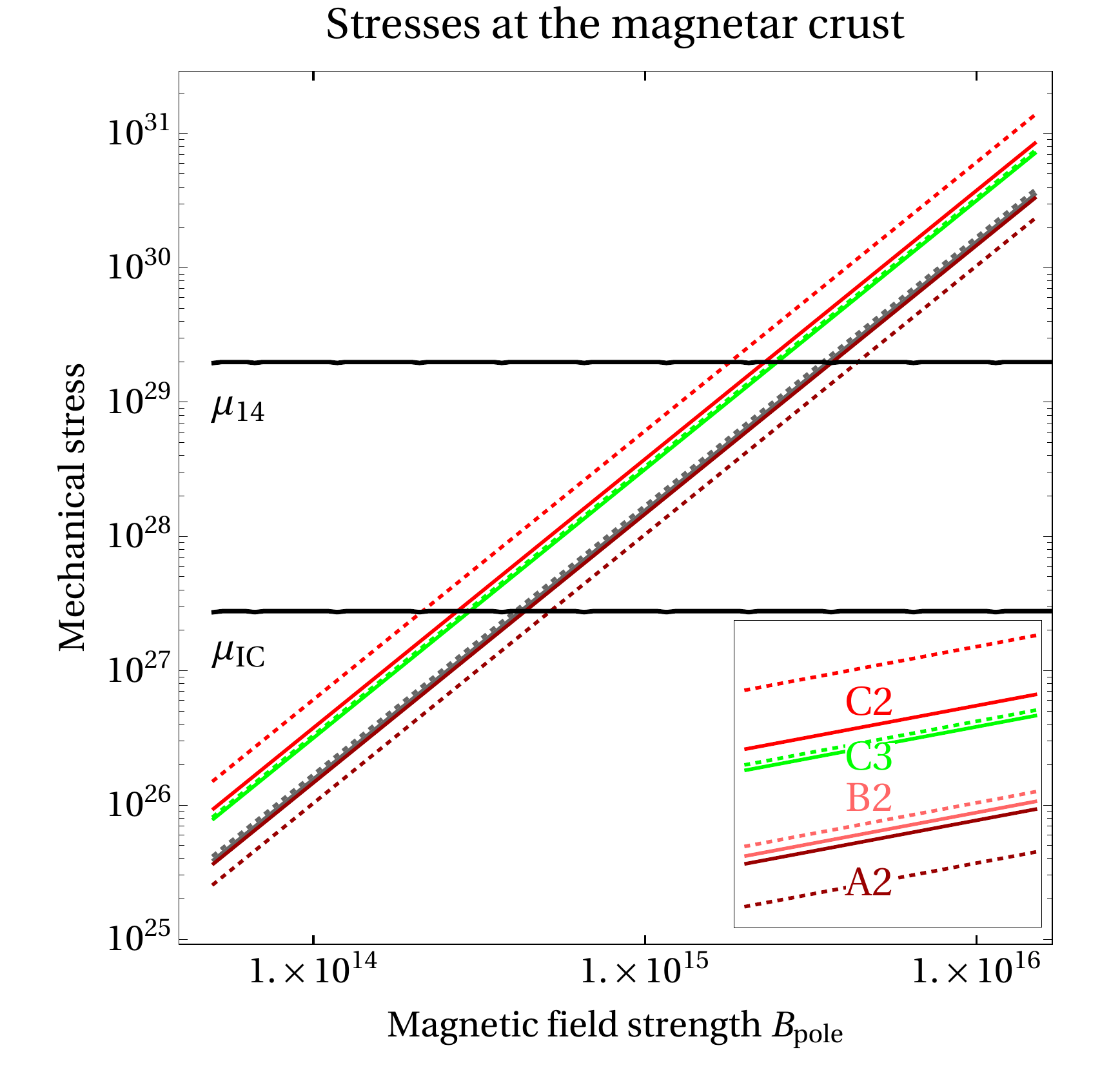} 
	\vspace{-11pt}
	\caption{Mechanical stresses exerted on the magnetar crust
 (according to equation~\ref{eq:conversion_stresses}) for the
 maximum stresses (Tables~\ref{tab:model_overview} and~\ref{tab:model_results}) observed during the high resolution
 simulations of models A2, B2, C2, and C3. The stress
 component $\sigma^{r\varphi}$ is denoted by solid lines, the
 component $\sigma^{r\theta}$ by dotted lines. The color
 coding corresponds to the initial models as introduced in
 Figure~\ref{fig:absolute_energy_evolution}. The black lines
 denote the approximate breaking stresses
 $\sim 0.2 \mu_{\rm IC}$, and $\sim 0.2 \mu_{\rm 14}$, at the
 inner crust boundary and near the core-crust transition,
 respectively. The high energy models reach the limit of a
 possible breaking of field lines for field strengths of
 $B_{\rm pole}\approx 10^{15}-10^{16}\text{ G}$.}
	\label{fig:shear_increase}
\end{figure}

For the discussion at hand, we will consider that the anchoring is
produced at some point between the inner crust outer boundary
($\rho\approx 4\times 10^{11}$\,g\,cm$^{-3}$), with
$\mu_{\rm IC}\approx 1.4 \times 10^{28}$\,erg\,cm$^{-3}$, and
$\mu_{14} \sim 10^{30}$\,erg\,cm$^{-3}$, its value close to the
core-crust transition, at about $10^{14}$\,g\,cm$^{-3}$.
The relevant components of the stress tensor in spherical
coordinates are
\begin{align}
\sigma^{r\theta} = 2\mu \, s^{r\theta}&= \mu\, \left [ r \partial_r \left (\frac{\xi^\theta}{r} \right) + \frac{1}{r} \partial_\theta \xi^r \right ] , \\
\sigma^{r\varphi}= 2\mu\, s^{r\varphi} &= \mu\, \left [ r \partial_r\left( \frac{\xi^\varphi}{r} \right )+ \frac{1}{r \sin\theta} \partial_\varphi \xi^r \right ],
\end{align}
where $s^{ij}$ is the strain tensor. For sufficiently large strains the crust will fail and a rapid plastic deformation will deform the crust persistently. 
The breaking strain of the crust has been estimated to be about $0.1$ \citep{Horowitz2009}. Therefore, any stress
larger than $\sim 0.2 \mu_{\rm c}$ will likely produce a failure in the crust. The maximum mechanical stress exerted on the magnetar crust, $\sigma^{ra}_{\rm max}$, can be quantified directly from the results shown in Tables~\ref{tab:model_overview}, and~\ref{tab:model_results} by employing the conversion formula 
\begin{align}
\sigma^{ra}_{\rm max}=5.55\times 10^{28} \,\text{erg}\,\text{cm}^{-3}\left ( \frac{\Delta_{\rm mx}T^{ra}}{0.1} \right )
 \left ( \frac{\TT^{ra}_{\rm max}}{10^{-9}} \right ) 
\left(\frac{B_{\rm pole}}{10^{15}\,\text{G}}\right)^2\!\!\!.
\label{eq:conversion_stresses}
\end{align}
The maximum mechanical stress (see Figure~\ref{fig:shear_increase}) on the magnetar crust measured throughout the shown simulations (see tabs. \ref{tab:model_overview} and \ref{tab:model_results}) correspond to $\sigma^{ra}\approx 10^{28}\,\text{erg}\,\text{cm}^{-3}$ for $B_{\rm pole}\approx 10^{15}\,$G. Considering the quadratic leverage of the magnetic field strength, mechanical stresses of $\sigma^{ra}\approx 10^{30}\,\text{erg}\,\text{cm}^{-3}$ are likely to be reached for $B_{\rm pole}\approx 10^{16}\,$G and beyond. The largest mechanical stresses are exerted in case of the high energy models A2, B2, C2, and C3. 

Our numerical simulations indicate that the instability occurs in a
quasi-axisymmetric way (cf. Figure~\ref{fig:high_energy_relaxation}),
with deviations from axisymmetry of less than
 $1\%$.\footnote{We quantify these deviations performing a
 multipolar expansion of the eletromagnetic energy and evaluating
 the energy stored in modes with azimuthal numbers $m>0$.} In
axisymmetry, axial displacements ($\xi^\varphi$) and polar
displacements ($\xi^r,\xi^\theta$) decouple and it is possible to
estimate the axial displacement from the $\sigma^{r\varphi}$ component
of the stress tensor. Although the magnetospheric dynamics can in
principle induce radial deformations, $\xi^r$, in reality those
deformations are strongly suppressed because they involve the motion
of matter parallel to the gravitational field (not included in our
calculation). Therefore, in practice one can consider $\xi^r=0$, such
that
\begin{align}
\sigma^{r\varphi} &= \mu\, r \partial_r\left( \frac{\xi^\varphi}{r}
 \right ), \\
\sigma^{r\theta} &= \mu\, r \partial_r\left( \frac{\xi^\theta}{r} \right ).
\end{align}
The transition at the anchoring point happens across a small distance, $h\equiv R_*-r_{\rm c}$, 
over which we can consider that $\mu=\mu_{\rm c}$ and $\sigma^{ra}$ are constant. Integrating the stress tensor along this distance we obtain:
\begin{align}
\xi^{a}_{\rm c} = r_{\rm c} \frac{\sigma^{ra}}{\mu_{\rm c}} \ln \left(\frac{r_{\rm c}}{r_{\rm c} + h}\right)
\approx R_* \frac{\sigma^{ra}}{\mu_{\rm c}}\label{eq:approx_displacement},
\end{align}
for $h\ll r_{\rm c},R_*$, and independent of the size of the transition
layer, $h$. The radial force per unit volume induced by the applied stress is \citep{Landau:elasticity}
\begin{align}
f^r = \sigma^{rk}_{\quad; k} =\frac{1}{r} \left ( \partial_\theta \sigma^{r\theta} + \cot\theta \sigma^{r\theta} \right )+ \frac{1}{r\sin\theta} \partial_\varphi \sigma^{r\varphi},
\end{align}
where we have considered that the only non-vanishing components are
$\sigma^{r\theta}$ and $\sigma^{r\varphi}$. We can estimate the
radial displacement $\xi^r$ balancing this force with the
gravitational force on the displaced mass, taken out of hydrostatic
equilibrium. We can make an order of magnitude estimate using
 linear perturbation theory if one neglects terms including gradients
 of background quantities and perturbations of the gravitational
 potential. In that case, the force balance reads:
\begin{align}
c_s^2 \,\rho \,\partial_{rr} \xi^r \approx -f_r.
\end{align}
Integrating over the transition length $h$ we get
\begin{align}
\xi^r \approx -\frac{f_r\,h^2}{2\,c^2_s\,\rho} \approx -\frac{c^2_{\rm shear}}{c_s^2} h^2 s^{rk}_{\quad;k},
\end{align}
where $c^2_{\rm shear} \equiv \mu/\rho$ is the shear speed. For typical values in the crust one assumes
$c^2_{\rm shear}/c^2_s \sim 10^{-2}$. If we consider the maximum
possible strain, i.e. the breaking strain,
$s^{ij} \sim h s^{ij}_{\:\:\: ;j}\sim 0.1$ \citep{Horowitz2009}, and the
maximum possible value for $h\sim\Delta R\sim 1$\,km, the size of the
crust, one finds an upper limit for the radial displacement of
$\xi^r_{\rm max} \sim 100$\,cm. At the same time, the displacement
components may be estimated directly from the results displayed in
Figure~\ref{fig:shear_increase} by employing
equation~(\ref{eq:approx_displacement}) and
$\mu_{\rm c}=0.5\times\left(\mu_{\rm 14}+\mu_{\rm IC}\right)$:
\begin{align}
\xi^{a}_{\rm c} &\approx 2.7\times10^{4}\,\text{cm}\,\left ( \frac{\sigma^{ra}}{10^{28}\,{\rm erg\,cm}^{-3}} \right )\left(\frac{B_{\rm pole}}{10^{15}\,\text{G}}\right)^2 
 \left( \frac{R_*}{\Rstar}\right )\,,
\end{align}
Our results show that for typical magnetar field strengths ($B\gtrsim 10^{15}$\,G)
the instability is likely to break a large fraction of the crust down to the inner crust. For the largest
magnetic fields ($B\gtrsim 10^{16}$\,G) the stresses induced in the crust are sufficient to shatter the 
entire crust. We should mention that the three magnetars that have showed GFs are among the more magnetised known ones and all three exceed $5\times 10^{14}$\,G.

\subsection{Emission processes}
\label{sec:emission_processes} 

%
\begin{table*}
	\centering
 \caption{Energetics of our models scaled to a
	magnetic field strength $B_{\rm pole}=10^{15}\,$G. (i) Energy released. (ii) Estimates of dynamic luminosity $L_0$ (equation~\ref{eq:L0}). (iii) Estimates of photospheric luminosity $L_{\rm ph}$
	(equation~\ref{eq:Lphotos}). (iv) Estimates of photospheric temperature $k_{\textsc{b}}T_{\rm ph}$
	(equation~\ref{eq:Tphotos}). Rows (v) and (vi) display
	the estimated photospheric luminosity $\mathcal{L}_{\rm ph}$
	and temperature $k_{\textsc{b}}\mathcal{T}_{\rm ph}$
	computed for the case in which $\eta < \eta_\ast$, assuming
	that the energy is released over a timescale
	$\Delta t_{\rm spike}=0.1\,$s (equation~\ref{eq:L0eff}). Finally,
	rows (vii) and (viii) show the initial luminosity
	$\mathcal{L}_0$ (equation~\ref{eq:L0eff}) and temperature
	$k_{\textsc{b}}\mathcal{T}_0$ also assuming that the energy is released over a timescale
	$\Delta t_{\rm spike}=0.1\,$s. Note that the last two rows
	coincide with the photospheric values if $\eta > \eta_\ast$.}
\label{tab:effective_temperature}
	{\renewcommand{\arraystretch}{1.5}
		\begin{tabular}{cccccc}
			\hline
			& & C2 & C3 & B2 & A2 \\\hline
			(i) & $\ee_{\rm r}$ (erg) & $6.03\times 10^{46}$ & $4.21\times 10^{46}$ & $2.15\times 10^{46}$ & $2.07\times 10^{46}$ \\
			(ii) & $L_0$ (erg\,s$^{-1}$) & $2.78\times 10^{49}$ & $4.00\times 10^{49}$ & $7.05\times 10^{48}$ & $1.46\times 10^{48}$ \\
			(iii) & $L_{\rm ph}$ (erg\,s$^{-1}$) & $9.32\times 10^{47}$ & $2.31\times 10^{47}$ & $5.9\times 10^{47}$ & $3.49\times 10^{47}$ \\
			(iv) & $k_{\textsc{b}}T_{\rm ph}$ (keV) & $25$ & $21$ & $43$ & $84$ \\\hline
			(v) & $\mathcal{L}_{\rm ph}$ (erg\,s$^{-1}$) & $2.60\times 10^{47}$ & $2.31\times 10^{47}$ & $1.84\times 10^{47}$ & $1.82\times 10^{47}$ \\
			(vi) & $k_{\textsc{b}}\mathcal{T}_{\rm ph}$ (keV) & $121$ & $140$ & $186$ & $189$ \\\hline
			(vii) & $\mathcal{L}_0$ (erg\,s$^{-1}$) & $6.03\times 10^{47}$ & $4.21\times 10^{47}$ & $2.15\times 10^{47}$ & $2.07\times 10^{47}$ \\
			(viii) & $k_{\textsc{b}}\mathcal{T}_0$ (keV) & $281$ & $257$ & $217$ & $215$ \\\hline
	\end{tabular}}
\end{table*}
\subsubsection{Estimation of observational properties of the
 energy release}
\label{sec:obsproperties}
\begin{figure}
	\centering
	\includegraphics[width=0.49\textwidth]{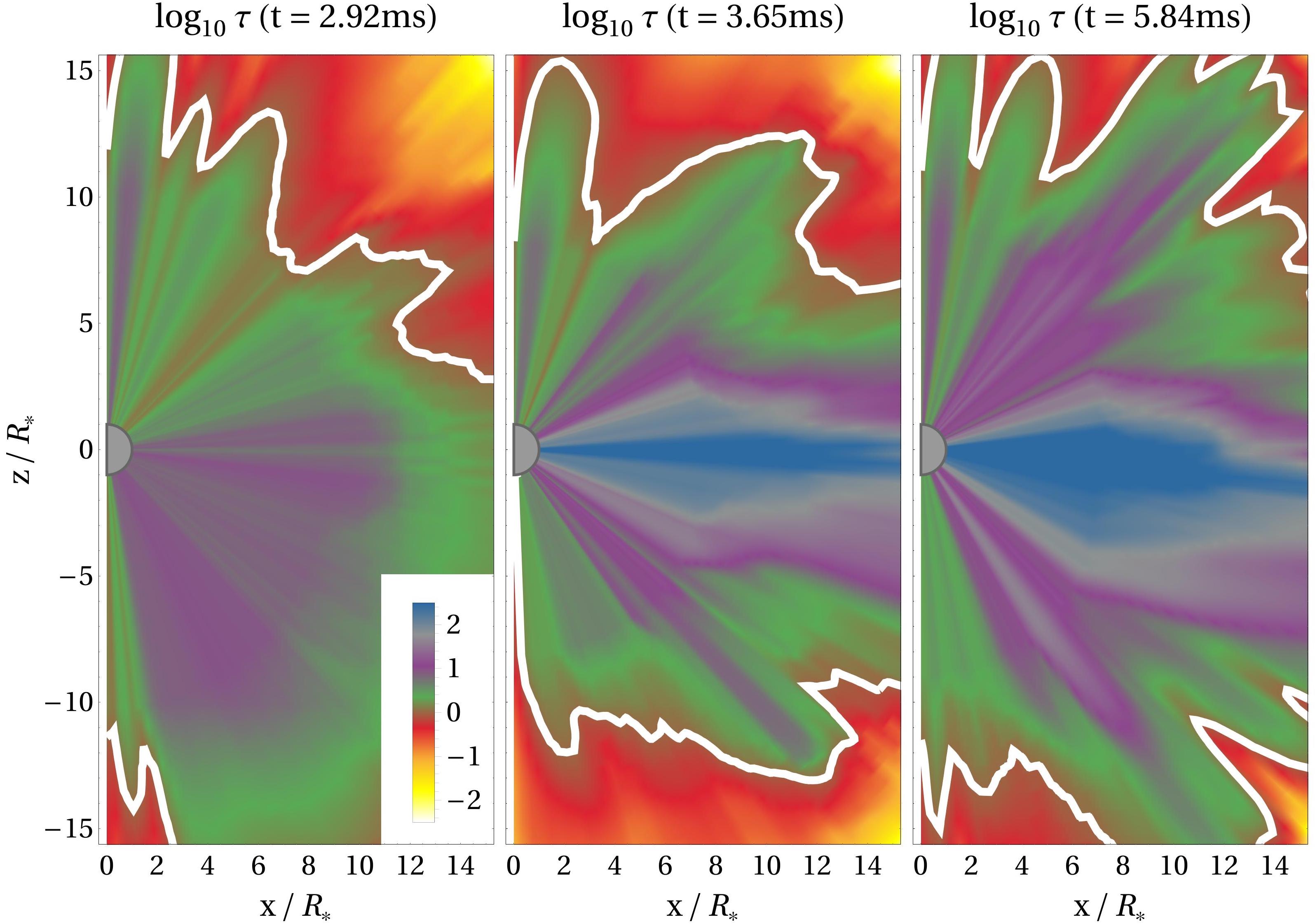} 
	\vspace{-11pt}
	\caption{Snapshots of the logarithm of the optical thickness
		during the evolution of the high resolution version of model
		C2. The logarithm of the optical thickness for the
		$\left\{\mathcal{M} = 100, \gamma=30\right\}$ model is
		displayed by the colour scale, the photosphere ($\tau=1$) is
		displayed as a white solid line. See appendix
		\ref{sec:optical_thickness} for further details.}
	\label{fig:optical_thickness_colourized}
\end{figure}
\begin{figure}
	\centering
	\includegraphics[width=0.45\textwidth]{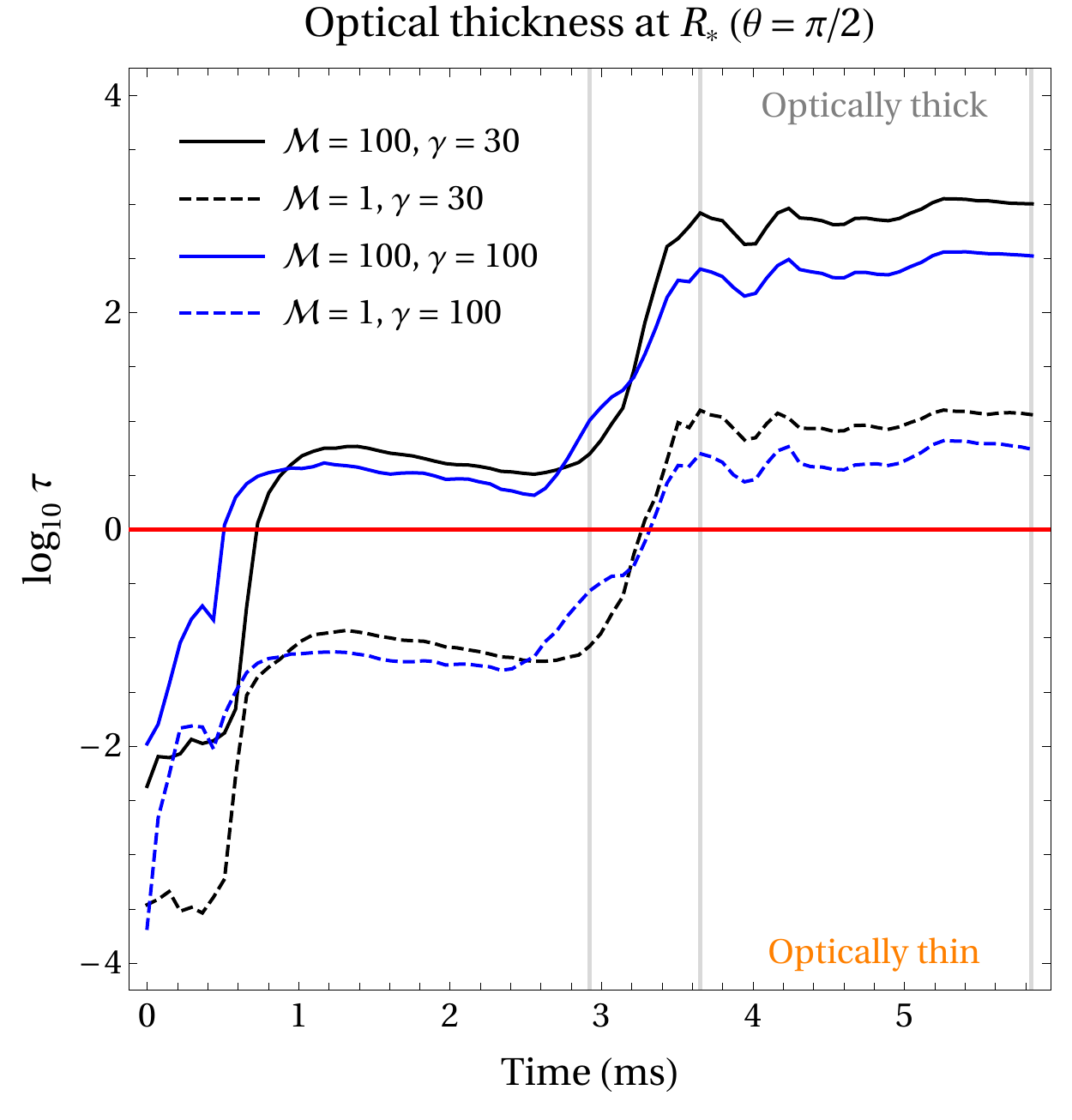}
	\vspace{-11pt} 
	\caption{Evolution of the optical thickness at the equator of the magnetar ($\theta=\pi/2$,
		$r=R_*$) of the high resolution initial model C2 ($B_{\rm pole}=10^{15}\,\text{G}$)
		and various parameter sets
		(see legends). The modeled optical thickness depends
		sensitively on the chosen multiplicity $\mathcal{M}$ and
		Lorentz factor $\gamma$ (cf. section~\ref{sec:optical_depth} and
		appendix \ref{sec:optical_thickness}). Initially, the
		magnetosphere is optically thin. During the onset of the
		instability ($\sim 1$\,ms) charges are produced in the
		magnetosphere. Depending on the chosen model ($\mathcal{M}$,
		$\gamma$), the magnetosphere becomes optically thick
		at different times. The times used for the visualisation of optical thickness in Figure~\ref{fig:optical_thickness_colourized} are denoted in gray lines.
	}
	\label{fig:optical_thickness_evolution}
\end{figure}

We have advanced that our models may release $\ee_{\rm r}\approx
2.1\times 10^{46}-6.4\times 10^{46}$\,erg on timescales of
milliseconds, producing, therefore, dynamic luminosities $L_0 \sim
(0.7 - 4) \times 10^{49}\,$erg\,s$^{-1}$ for the unstable models
listed in Table~\ref{tab:effective_temperature}. Following the
reasoning of \cite{Thompson1995}, confining this energy in
the form of a photon-pair plasma by a closed magnetic flux loop of
outer radius $r$ requires that the field pressure at the
outer boundary of the loop exceed the deposited energy density
 \begin{align}
 \frac{B(r)^2}{8\pi}\gtrsim
 \frac{\ee_{\rm r}}{4\pi/3 (r^3-R^3)}.
\end{align}
In a dominantly dipolar magnetosphere, $B(r)\sim
B_{\rm pole}(R_*/r)^3$, the plasma can be confined if $\ee_{\rm r} < \ee_{\rm d}/2$ 
within a radius $r$ in the range
\begin{align}
 \frac{\ee_{\rm d}}{\ee_{\rm r}} \left[ 1 - \sqrt{1 - 2 \frac{\ee_{\rm r}}{\ee_{\rm d}} } \right ]
\lesssim \left ( \frac{r}{R_*} \right )^3 \lesssim
 \frac{\ee_{\rm d}}{\ee_{\rm r}} \left[ 1 + \sqrt{1 - 2 \frac{\ee_{\rm r}}{\ee_{\rm d}} } \right ].
\end{align}
For the range of values of $\ee_{\rm r}/\ee_{\rm d} \sim \Delta \ee/\ee_{\rm
 d}$ from our models (Table~\ref{tab:model_results}) we obtain that the 
 size of the confinement region, $\Delta R\equiv r-R_*$, is limited by 
\begin{align}
(1.8-6)\times 10^{-2}\, R_* \lesssim \Delta R \lesssim (0.8-1.7)\, R_*.
\label{eq:DeltaR}
\end{align}
Note that this result is independent of the magnetic field strength $B_{\rm pole}$.

Our numerical simulations show that most of the energy is released in
a thin and numerically unresolved surface current of the
star, that we measure as a Poynting flux (see
Figure~\ref{fig:energy_dissipation_channels}) and in a region close to
the surface ($r\lesssim1.25 R_*$) with large currents (see
Figures~\ref{fig:surface_currents_xz} and~\ref{fig:surface_currents_avg}). Energy deposited there, essentially
at the footprints of magnetic field lines, is expected to distribute
efficiently along those lines aided by the flowing pair plasma. As a
result, we expect that the energy will fill an extended region of the
magnetosphere comparable in size to the region filled with currents
(see Figure~\ref{fig:high_energy_relaxation}). This region can be as
large as $\sim 4 R_*$ at the time of maximum energy dissipation. For
magnetic field lines extending within the limits given by
equation~(\ref{eq:DeltaR}), the energy is expected to be
confined. However, for lines extending beyond $(0.8-1.7) R_*$, the
energy will not be confined and it may yield an ultrarelativistic
fireball composed of pairs, photons and a small amount of baryons
lifted up from the outer crust by the large energy released
there. Depending on the structure of the magnetosphere, the energy
released in this form can be a significant fraction of $\ee_{\rm r}$.
Obviously, our methodology does not allow us to track the evolution of
the released energy, but we may obtain a rough estimation of its
bolometric properties. For the estimate we will consider that most of
the energy is released in the fireball, which gives us upper limits.

The physics of such expanding fireball has been considered in many
papers
\citep[e.g.][]{Goodman_1986ApJ...308L..47,Paczynski_1986ApJ...308L..43,
 Shemi_1990ApJ...365L..55, Piran_1993MNRAS.263..861,
 Meszaros_1993ApJ...415..181}, especially addressing the generation
of gamma-ray burst (GRBs), but also applied to SGRs
\citep[e.g.][]{Nakar_2005ApJ...635..516}. The sudden energy release
results into a a thermal burst carrying {\it most} of the initial
energy, and according to the cannonical interpretation
\citep[e.g.][]{Hurley2005}, with roughly the original temperature and
a fraction of the energy in the form of relativistic pairs. The
observed thermal spectrum of the flare and its temperature support
this idea.

Here we follow the model of
\cite{Meszaros:2000ApJ...530..292}, which suffices for the basic
estimates we aim at. Assuming that in a region of
size $R_0\simeq R_*$ (initially at rest), energy is released at a rate $L_0$, the initial
temperature of the fireball in units of the electron rest mass
is \citep[][equation~2]{Meszaros:2000ApJ...530..292}
\begin{align}
\Theta_0 &= \left(\frac{k_{\textsc{b}}}{m_e c^2}\right)\left( \frac{L_0}{4\pi
 R_0^2 c a_r}\right)^{1/4} \nonumber \\& =1.43\,
\left( \frac{L_0}{\Lo}\right)^{1/4}
\left( \frac{R_0}{\Rstars} \right)^{-1/2},
\label{eq:Theta0}
\end{align}
where $m_e=9.1095\times 10^{-28}\,$g is the electron mass, $a_r=7.57\times 10^{-15}\,$g$\,$cm$^{-1}\,$s$^{-2}\,$K$^{-4}$
is the radiation constant and
$k_{\textsc{b}}\simeq 1.38\times 10^{-16}\,$erg\,K$^{-1}$ is the
Boltzmann constant. In the previous equation (and hereafter) we have
scaled the luminosity to the dynamical luminosity estimated for model
C2, but a similar exercise has been undertaken for models C3, B2 and
A2, being the results listed in
Table~\ref{tab:effective_temperature}.
 The value of $\Theta_0$ in
equation~(\ref{eq:Theta0}) corresponds to a comoving temperature
$k_{\textsc{b}}T_0\simeq 732\,$keV. 
Starting from its initial radius, $R_0$, the
fireball expands and accelerates until it converts most of its
internal energy into kinetic energy at a distance $R_{\rm s}$,
commonly called the saturation radius (see equations~\ref{eq:Rsat} and
\ref{eq:Rsat2} below). The Lorentz factor, $\Gamma$, of the expanding
fireball is approximately given by
\begin{align}
\Gamma = \begin{cases} \displaystyle\frac{r}{R_0} & \text{if}\:\: r<R_{\rm s},
\\*[5pt]
\displaystyle\frac{R_{\rm s}}{R_0} & \text{if}\:\: r\ge R_{\rm s}. 
\end{cases}
\end{align}
The amount of mass that may be unbound due to an energy release as large as suggested by our models ($\ee_{\rm r}$) is uncertain, but we may estimate it to be as small as $M_{\rm ex}\simeq 3\times 10^{-10}M_\odot$. The period over which this mass is extracted we
assume to be the same as that over which the energy is released, $\Delta
t_{\rm r}$. This implies a mass loss rate from the magnetar surface
$\dot{M}\simeq M_{\rm ex}/\Delta t_{\rm r} \simeq 2.8\times
10^{26}\,$g\,s$^{-1}$. The dimensionless entropy of the fireball for
this baryon load is
\begin{align}
 \eta &=\frac{L_0}{\dot{M}c^2} \nonumber \\ & \simeq \etao\, 
\left( \frac{L_0}{\Lo} \right) 
\left( \frac{ \dot{M} }{2.8 \times 10^{26}\,\text{g\,s}^{-1}} \right)^{-1}. 
\label{eq:eta0}
\end{align}
As usual, we define the photospheric radius as the distance at which
the fireball becomes optically thin, which may happen before the
Lorentz factor saturates or after that, i.e. in the regime where the
fireball coasts
\begin{align}
R_{\rm ph} & \simeq \frac{L_0\sigma_{\textsc{t}}Y}{4\pi m_{\rm p}c^3
 \eta^3}, \label{eq:Rphotos} \qquad\qquad (R_{\rm ph} > R_{\rm s}) \\
R_{\rm ph} &\simeq \left(\frac{L_0\sigma_{\textsc{t}}Y}{4\pi m_{\rm p}c^3
 \eta}\right)^{1/3}. \qquad (R_{\rm ph} \le R_{\rm s})
\label{eq:Rphotos2}
\end{align}
Here, $\sigma_{\textsc{t}}=6.6525\times 10^{-25}\,$cm$^2$ and
$m_{\rm p}=1.6726\times 10^{-24}\,$g are the Thompson cross-section
and the proton mass, respectively. $Y$ represents the number of
electrons per baryon. In the following, we will take $Y\simeq 1$,
which is appropriate once pairs are not present in the system. Indeed,
this shall be the case for radii larger than $R_{\rm p}$
\citep[][equation~3]{Meszaros:2000ApJ...530..292}
\begin{align}
\begin{split}
R_{\rm p}=\:& R_0 \frac{\Theta_0}{\Theta_p} \simeq 5.8\times 10^{7}\,\text{cm} \\
&\times\left( \frac{L_0}{\Lo}\right)^{1/4} 
\left( \frac{R_0}{\Rstars} \right)^{1/2} \left( \frac{\Theta_p}{0.03}\right)^{-1},
\end{split}
\end{align}
where the comoving dimensionless temperature below which $e^\pm$ pairs
drop out of equilibrium is $\Theta_p\simeq 0.03$ (equivalently,
$k_{\textsc{b}}T_p\simeq 17.4\,$keV). Note that $R_p\ll R_{\rm ph}$
(see equations~\ref{eq:Rphotos} and \ref{eq:Rphotos2}).

The critical baryon load, $\eta_\ast$ for which the photospheric
radius equals the saturation radius, i.e. $R_{\rm ph} = R_{\rm s}$,
is given by \citep[][equation~5]{Meszaros:2000ApJ...530..292}
\begin{align}
\begin{split}
\eta_\ast &= \left( \frac{L_0 \sigma_{\textsc{t}}}{4\pi m_p c^3 R_0}
\right)^{1/4} \\
 &\simeq \etaaster \,
\left( \frac{L_0}{\Lo}\right)^{1/4}
\left( \frac{R_0}{\Rstars} \right)^{-1/4}.
\end{split}
\label{eq:eta-star}
\end{align}
Depending on the value of the parameter $\eta$, there are two
regimes. Either the photospheric radius happens beyond the saturation
radius ($\eta < \eta_\ast$) or, otherwise, the saturarion radius
happens when the fireball is still expanding ($\eta > \eta_\ast$). In
the former case, the saturation radius is
\begin{equation}
R_{\rm s} =\eta R_0 \simeq 1.5\times 10^8\,\text{cm}\,
\left( \frac{R_0}{\Rstars}\right)
\left(\frac{\eta}{\etao}\right)\,,
\label{eq:Rsat}
\end{equation}
where we have used for $\eta$ the value computed in equation~(\ref{eq:eta0}) for
the assumed value of $\dot{M}$. If the photosphere appears when the fireball is still
accelerating, the saturation radius is attained at a distance \citep[][equation~11]{Meszaros:2000ApJ...530..292}
\begin{align}
R_{\rm s} =\:& \eta_\ast R_0 \nonumber \simeq 5.4\times 10^{8}\,\text{cm} \\
&\times\left( \frac{L_0}{\Lo}\right)^{1/4}
\left( \frac{R_0}{\Rstars} \right)^{3/4},
\label{eq:Rsat2}
\end{align}
Interestingly, \cite{Hurley2005} model the peak of SGR\,1806-20
assuming that the dimensionless entropy of the fireball is $\eta
>\eta_\ast$ because for the observed peak luminosity (much smaller
than that implied in our models, namely, $\sim 2\times
10^{47}\,$erg\,s$^{-1}$), the critical baryon load would be $3-4$
times smaller than estimated in equation~(\ref{eq:eta-star}) and, hence,
\citet{Hurley2005} naturally obtain $\eta \gtrsim
\eta_\ast^{(1806-20)}$. The observational difference between the two
described regimes is notable for our models as we see next in the
estimation of the photospheric temperature and luminosity of the events.
In the case $\eta<\eta_\ast$, the photospheric temperature and
luminosity are, respectively,
\begin{align}
\begin{split}
k_{\textsc{b}}T_{\rm ph} =k_{\textsc{b}} T_0 \left(\frac{R_{\rm ph}}{R_{\rm s}}\right)^{-2/3} \simeq\:&
25\,\text{keV}\,
\left( \frac{L_0}{\Lo}\right)^{-5/12}\\
&\times
\left(\frac{\eta}{\etao}\right)^{8/3} \left( \frac{R_0}{\Rstars} \right)^{-5/6},
\end{split}
\label{eq:Tphotos}
\end{align}
and \citep[][equation~9]{Meszaros:2000ApJ...530..292}, 
\begin{align}
\begin{split}
L_{\rm ph} =& L_0 \left(\frac{R_{\rm ph}}{R_{\rm s}}\right)^{-2/3}\\ \simeq\:&
9.3\times 10^{47}\,\text{erg}\,\text{s}^{-1}\,
\left( \frac{L_0}{\Lo}\right)^{7/12}\\
&\times
\left(\frac{\eta}{\etao}\right)^{8/3} \left( \frac{R_0}{\Rstars} \right)^{-5/6}.
\label{eq:Lphotos}
\end{split}
\end{align}

The value obtained in equation~(\ref{eq:Tphotos}) must be compared with the
ones obtained from observations, namely $k_{\textsc{b}}T^{\rm obs}_{\rm peak} \simeq
175-250\,$keV. Our result underestimates the observed temperature
significantly. However, we are neglecting Comptonisation effects,
which may slightly raise the estimated photospheric temperature (still
below the observational data). Note that smaller values of $L_0$,
in line with the observed luminosities at peak for SGRs, would bring
the observed photospheric temperature to the observed values, but, at
the same time, they would significantly raise the photospheric
luminosity, hence yielding events much more luminous than
observed. The dependence on $\eta^{8/3}$ is the same in both equations (\ref{eq:Tphotos}) and (\ref{eq:Lphotos}), therefore, changes in the
assumed baryon loading may not improve the consistency of our
estimated photospheric values with the observed ones. However, if the
baryon load is sufficiently small such that $\eta>\eta_\ast$ \citep[as
assumed in][]{Hurley2005}, the
declining temperature and luminosity in the outflow are compensated by the
relativistic blueshift. In this case, we would estimate the following
photospheric temperature
\begin{align}
k_{\textsc{b}}T_{\rm ph} &= k_{\textsc{b}} T_0 \nonumber \\
&\simeq
723\,\text{keV}\,
\left( \frac{L_0}{\Lo}\right)^{1/4}
\left( \frac{R_0}{\Rstars} \right)^{-1/2},
\end{align}
and luminosity
\begin{align}
L_{\rm ph} &= L_0 \simeq \Lo\,\text{erg}\,\text{s}^{-1}.
\end{align}
In this case, both estimations for $T_{\rm ph}$ and $L_{\rm ph}$ significantly
overestimate the observed values for SGRs.

We have found in this section that independently of whether the
photosphere of the expanding fireball happens in the acceleration
phase or in the coasting phase, the estimated values of $T_{\rm ph}$
and $L_{\rm ph}$ are not compatible with observations. The root for
the discrepances found are the very large dynamic luminosities ($L_0$)
of most of our models. These large values result from considering
magnetospheric initial data where the twist is so large that they
release a large amount of energy on timescales of milliseconds. We
note that models with larger relative toroidal fields (as induced by a
power-index $\sigma=2$, and $s=2$) spaning a larger fraction of the
magnetar surface (due to their smaller values of $P_{\rm c}$), e.g. model A2
(Table~\ref{tab:effective_temperature}), show values of $T_{\rm ph}$
and $L_{\rm ph}$ broadly compatible with the most energetic GFs
observed so far \citep[see, e.g.][]{Hurley2005,CotiZelati2018}. This
is in contrast to models where we have built up the magnetosphere with
$s=\sigma=1$ (namely, C2, C3 and B2), which systematically yield
over-luminous and too cold photosperic conditions. Thus, our results suggest that twisting magnetospheres to the largest (theoretical) levels we have considered
here may not be realised in nature. Well before reaching the largest
twists of models C3, C2 or B2 the dynamical instability may set in releasing
smaller amounts of energy (and hence, producing smaller dynamical
luminosities). 

A potential handicap in our models is the duration of the
observational signal that yield the fireballs modelled so far. In the
canonical fireball model, the energy release leads to a {\it frozen
 pulse} whose duration approximately equals the timescale over which
the energy is deposited, $\Delta t_{\rm r}$ \citep[e.g.][but see
\citealt{Janka_2006ApJ...645.1305}]{Piran_1993MNRAS.263..861}. Since
$\Delta t_{\rm r}\ll \Delta t_{\rm spike}$, the quasi-thermal
radiation bursts that we have estimated are too short to account for
the typical timescale of the initial spike of GFs in SGRs
($\Delta t_{\rm spike}\sim 0.1\,$s). In our simulations, the energy change in the magnetosphere is driven by the Poynting flux through the star surface. However, the ability of the crust to absorb all this energy on the dynamical timescale of the
magnetosphere is limited because of the low transmission coefficient 
(see equation~\ref{eq:transmission}).
So far we have considered that all this energy is temporarily stored in a thin layer above
the magnetospheric surface, where intense currents may convert
the stored magnetic energy into thermal energy. This is consistent with the boundary 
conditions imposed in our numerical simulations. Alternatively, we could have chosen
boundary conditions that avoid the formation of strong thin surface currents \citep[as e.g. in][]{Carrasco2016}.
In that case, Alfvén waves propagating towards the surface of the star get reflected and collide 
at some distance from the surface. This forces the formation of reconnection points at some distance from the neutron star surface. \cite{Li2018} have estimated that this process is relatively inefficient in 
dissipating the energy of the magnetosphere and that it may take multiple bounces in the 
magnetosphere to dissipate all the energy. This may allow for a slower energy deposition
on timescales $\sim \Delta t_{\rm spike}$. 

Unfortunately, our numerical
models do not include the relevant microphysics to fully address the
conversion of magnetic into thermal energy. Thus, we can only warn the
reader that the milliseconds timescales over which we have made our
(simple) estimations of the dynamical luminosity of the models at hand
are only lower bounds of the true timescales on which the released
energy may leave the magnetosphere. Taking into account this caveat,
the values of $L_0$ listed in Table~\ref{tab:effective_temperature}
are upper bounds to the effective initial luminosity, $\mathcal{L}_0$,
\begin{equation}
 \mathcal{L}_0:=\frac{\ee_{\rm r}}{\Delta t_{\rm spike}}\simeq
 10^{47}\,\text{erg}\,\text{s}^{-1} \, \left(\frac{\ee_{\rm
 r}}{10^{46}\,\text{erg}}\right) \left(\frac{\Delta t_{\rm spike}}{0.1\,\text{s}}\right)^{-1} .
 \label{eq:L0eff}
\end{equation} 
Redoing the previous estimations for the photospheric conditions, we
find the values $\mathcal{L}_{\rm ph}$ and
$k_{\textsc{b}}\mathcal{T}_{\rm ph}$ listed in
Table~\ref{tab:effective_temperature}. In addition to these estimates
of the photospheric luminosity and temperature corresponding to the
values of the initial luminosity given by equation~(\ref{eq:L0eff}) when the
photosphere happens beyond the saturation radius (i.e. for
$\eta < \eta_\ast$), we also provide the estimation of the
photospheric luminosity ($\mathcal{L}_0$) and temperature
($k_{\textsc{b}}\mathcal{T}_0$) in the complementary case when the
photospheric conditions are reached during the acceleration phase of
the fireball (i.e. $\eta > \eta_\ast$). All these new values of the
photospheric luminosity and temperature are perfectly compatible with
observational data. Not surprisingly, we find that depending on
whether we assume that photospheric conditions are met in the
accelerating phase or in the coasting phase of the fireball, the
values obtained for the photospheric temperature {\it bracket} the
typical values found for the spike of SGRs.
\subsubsection{Optical depth of the magnetosphere}
\label{sec:optical_depth} 

The observed maximum current density throughout the magnetosphere, $J_{\rm max}$, can be quantified directly from the results shown in Tables~\ref{tab:model_overview}, and~\ref{tab:model_results} by employing the conversion formula
\begin{align}
\begin{split}
J_{\rm max}=4.4926\times
10^{12}\,\text{A/m}^2\,\left ( \frac{\JJ_{\rm max}}{10^{-6}} \right )\left(\frac{B_{\rm
		pole}}{10^{15}\,\text{G}}\right)\,.
\end{split}
\end{align}
The presented results compare well to the expected current density stated in equation~(\ref{eq:GJcurrent}). Close to the surface of the star, where the highest currents appear, the particle density is 
\begin{align}
n_e = \frac{J}{c e} \mathcal{M} \sim 10^{19}\,{\rm cm}^{-3},
\end{align}
where $\mathcal M$ is the multiplicity. \citet{Beloborodov2013} has estimated that in extended regions close to the poles the multiplicity 
can be as large as $\mathcal M \sim 100$, while close to the equator
$\mathcal M \sim 1$. 

The dominant contribution to the opacity in the magnetosphere
is the resonant cyclotron scattering of thermal photons off
charge particles in the vicinity of the neutron
star\footnote{If there is a dynamical mass ejection a result of the large energy release close to the magnetar surface (section~\ref{sec:obsproperties}), the Thompson scattering (in the
 expanding fireball) may be the dominant source of opacity at
 sufficiently large distances.\label{foot:one}}. \citet{Thompson2002} have estimated
that for twists of $\Delta\varphi \sim 1$ the typical optical depth in
the magnetosphere is $\sim 1$. In general, computing the optical depth
for magnetar magnetospheres is a complicated problem, because one
needs a self-consistent solution of the photon field and the momentum
distribution of charged particles traveling along the magnetic field
lines \citep[see][]{Beloborodov2013}. In this work we make an
estimation for radially streaming photons and a simplified momentum
distribution of charged particles. We only consider $1$\,keV photons,
which are typical for the observed surface temperature in
magnetars. Inspired by \citet{Beloborodov2013} we use a simple
waterbag momentum distribution (see
appendix \ref{sec:optical_thickness}) which is characterised by two
parameters, the mean specific momentum ($\bar p$, where $p=v W$) and
 $\mathcal{M}$. We integrate the optical depth
($\tau$) radially inwards (see app \ref{sec:optical_thickness},
equation~\ref{eq:optical_thickness_integrand} for details on the
computation) and identify the photosphere as the place where $\tau=1$.

Figure~\ref{fig:optical_thickness_colourized} shows estimates for the
optical thickness of the magnetosphere at three different times
(during and after the rapid drop of magnetospheric
energy) computed with parameters $\{\mathcal{M}=100, \gamma=30\}$. 
During the rearrangement of the magnetosphere, the coronal
region along the equator becomes optically thick. The initial configuration is optically thin and, hence, not shown here. An important conclusion is that close to the critical point, most of the magnetosphere, if not
all, is optically thin, which gives rise to a black body spectrum with
the typical temperature of the NS surface ($\sim 1\,$keV) plus a
possible non-thermal contribution of up-scattered photons. However,
during the instability, the increase of the magnetospheric currents,
makes a large fraction of the magnetosphere of a few stellar radii
optically thick. This region is filled up with pair plasma and will
emit thermal radiation through its photosphere. Its lifetime is
related to the presence of strong currents in the magnetosphere and
may be an explanation for the X-ray tail ($k_{\textsc{b}}T\sim30$\,keV) observed
after GFs and lasting for a few $100$\,s. We note
 that only a relatively small fraction of the total energy released
 in the magnetosphere by the instability may contribute to the tail,
while most of it may contribute to the initial peak characteristic of
GFs (see discussion in section~\ref{sec:obsproperties}). 

Our model to compute the magnetospheric optical thickness for
 resonant cyclotron scattering assumes uniform values of the
 multiplicity and of the electron Lorentz factor. Neither for the
 multiplicity (as we have argued above) nor for $\gamma$ this is
 completely correct. Modeling locally the values of the parameters
 $\{\mathcal{M}, \gamma\}$ is beyond the scope of this
 paper. However, we may test the robustness of our results by
 exploring the parameter space determined by $\mathcal{M}$ and
 $\gamma$. In Figure~\ref{fig:optical_thickness_evolution}, we display
 the time evolution of the optical thickness at the equator of the
 magnetar for various parameter sets. As expected, the larger the
 value of $\mathcal{M}$, the larger the number density of leptons
 and, consistently, the larger the opacity (note the nearly two
 orders of magnitude difference between the solid lines with
 $\mathcal{M}=100$ and the dashed lines with $\mathcal{M}=1$). The
 effect of the variation of the Lorentz factor (electrons or positrons) is
 small compared to the strong impact of $\mathcal{M}$ on the
 opacity. Although the magnetosphere becomes eventually optically
 thick for all the parameter sets under investigation, models with
 $\mathcal{M}=100$ develop regions with $\tau>1$ very early
 ($t\lesssim 0.8\,$ms), while models computed with $\mathcal{M}=1$
 become optically thick only when the instability in the
 magnetosphere fully develops.

 Emission by resonant scattering in
magnetar magnetospheres may be subject to ($\perp$ or $\parallel$)
polarisation \citep[see, e.g.][]{Fernandez2011,Beloborodov2013}. In
the presented (approximate) modeling of optical
thickness, however, we have found differences in these polarisation
states of $<1\%$. We will further explore the emission properties of
force-free twisted magnetospheres on suitable high-resolution
numerical data in our future work.

\section{Conclusions}
\label{sec:conclusion} 

In this work, we explore the stability properties of force-free equilibrium configurations of magnetar magnetospheres by performing numerical simulations of a selection of the models computed in \citet{Akgun2018a}. For the case of degenerate magnetospheres (i.e. the same boundary conditions but different energies) we validate the hypothesis of \citet{Akgun2018a} that configurations in the high-energy branches are unstable while those in the lowest energy branch are stable. This confirms the existence of an unstable branch of twisted magnetospheres. It also allows to formulate an instability criterion for the sequences of models computed in \citet{Akgun2018a}. Our results are consistent with an interesting scenario where bursts and GFs in magnetars are triggered without involving crustal failures. The twist that is naturally produced in the magnetosphere by the Hall evolution of the crust \citep{Akgun2017} can lead to unstable configurations that will release up to a 10\% of the energy stored in the magnetosphere, sufficient to explain the observations.

\cite{Akgun2017} have shown that the magneto-thermal evolution of the crust leads naturally to configurations close to the instability threshold. However, the amount of energy released depends on
how far away from the stable branch can the evolution drive the configuration. This is essentially a problem of comparing the evolution timescale and the instability timescale. For the models studied in 
this work the instability timescale is of the order of milliseconds, much shorter that the magneto-thermal evolution timescales of the object (see Sect.~\ref{sec:secular}). However, close to the critical point, 
the growth rate of the instability could be significantly smaller (actually, it should be zero at the critical point) which would allow to overshoot the instability threshold. Note that, since the energy reservoir
is large ($\sim 10^{46}$~erg), even a very small fraction of energy release could explain many of the phenomenology of magnetars. Alternatively, there could be phenomena leading to fast dynamics 
in the crust such as sustained episodes of accelerated plastic flows triggered by the magnetic stresses in the crust \cite{Lander2019}.

For the unstable models, we observe the development of almost axisymmetric instabilities on a timescale of a few ms rearranging the magnetic field to a configuration similar to those in the (stable) lower energy branch. The energy of the magnetosphere also decreases towards the value of the stable configuration. Differences with respect to the corresponding stable configuration can be attributed to the influence of the non-preservation of the force-free constraints (\ref{eq:force_free_crossfield}) and (\ref{eq:force_free_dominance}). Using (much) larger numerical resolution (beyond the scope of our computational resources) we envision that the violation of the force-free constraints would be significantly reduced and the expected (low-energy) states would be the endpoint of the evolution after a full relaxation of the magnetosphere takes place. The energy decrease is explained, mainly, by a flow of energy towards the surface of the star, where it is dissipated efficiently. A large fraction of this energy is also dissipated in the magnetosphere at locations where the force-free conditions break. This contrasts with the work of \citet{Beloborodov2011}, \citet{Parfrey2013} and \cite{Carrasco2019} in which most of the energy is dissipated by the formation and ejection of plasmoids. The different setup used in these workst (dynamically twisting vs. unstable equilibrium configurations) makes a direct comparison difficult. A possible source for the qualitative discrepancy may be differences in the boundary condition at the surface of the star. While we use a boundary condition that dissipates very efficiently any strong currents formed at the surface, in their work, their use of essentially non-dissipative boundary conditions make the surface perfectly reflective. For the future it would be interesting to compare more closely the differences in the boundary condition and to develop a better physical model for dissipation at the NS surface.

The magnetic field remains nearly axisymmetric throughout the simulation indicating that the instability is mostly an m=0 instability. A complete theoretical analysis of the origin of the instability and its properties is beyond the scope of this paper. However, we anticipate that such analysis has to be carried out on a global scale either by calculating the eigenmodes or by using the so-called energy principle of \citet{Bernstein1958} and is not trivial due to the presence of both poloidal and toroidal components \citep[][and references therein]{Akguen2013}. However, we note that, since the poloidal field structure changes somewhat less than the toroidal field, this instability could be compared to the interchange instability discussed by \citet{Tayler1973}, where displacing the toroidal field radially decreases the energy (even in the absence of a fluid).

 We have made a crude estimation of the observational properties of the energy liberated in the magnetosphere as a result of the instability. The fact that large amounts of energy (in excess of $10^{46}\,$erg) are released on milliseconds timescales results in dynamical luminosities significantly larger than $10^{48}\,$erg\,s$^{-1}$ (reaching in some models $4\times 10^{49}\,$erg\,s$^{-1}$). This should trigger the expansion of a pair-photon fireball polluted with baryons unbound from the magnetar crust. The bolometric signature of these fireballs seems incompatible with the observations of the initial spikes observed in GFs. With our simple analytic model, most of the unstable magnetospheres produce over-luminous, too cool and excessively short flashes. However, this problem can be solved if the energy can be liberated on longer timescales, of the order of the observed GF spikes ($\Delta t_{\rm spike}\sim 0.1\,$s). This could be possible in a scenario of slow energy dissipation as the one proposed by \cite{Li2018}, which we plan to explore in the future.

The currents produced during the instability increase significantly the amount of pairs in the magnetosphere, a large fraction of which, of size $\sim 10 R_*$, becomes optically thick. The hot plasma magnetically confined in this region could be responsible for the extended thermal X-ray emission lasting for $50-300$\,s after GFs.

Our force-free numerical method cannot properly deal with the evolution of extremely thin surface currents. Therefore, the dynamical millisecond timescales computed in our models should be taken as a lower bound for the physical timescales. The magnetic dissipation taking place at these locations can be due to, e.g. Ohmic processes or to non-linear Alfv\'en wave interactions. Assuming that energy is released on $\sim\Delta t_{\rm spike}$, our estimate of the electromagnetic signature yields photospheric luminosities and temperatures compatible with observational data. Since this is a sound physical assumption, we conclude that observed GFs in SGRs are broadly compatible with the development of instabilities in twisted magnetospheres.

\section{Acknowledgements}
We thank Amir Levinson for his support in challenging our numerical code prior to the production of the presented results. We also thank Oscar Reula, Federico Carrasco and Carlos Palenzuela for their valuable discussions on the boundary conditions. We acknowledge the support from the grants AYA2015-66899-C2-1-P and PROMETEO-II-2014-069. JM acknowledges a Ph.D. grant of the \textit{Studienstiftung des Deutschen Volkes}. PC acknowledges the Ramon y Cajal funding (RYC-2015-19074) supporting his research. We acknowledge the partial suport of the PHAROS COST Action CA16214 and GWverse COST Action CA16104. The shown numerical simulations have been conducted on \textit{MareNostrum 4} of the \textit{Red Espa\~{n}ola de Supercomputación} (AECT-2019-1-0004) as well as on the computational infrastructure of the \textit{University of Valencia}. We thank the \textit{EWASS 2019} and \textit{GR22} conferences for the possibility to disseminate the results of this work.

\bibliographystyle{mnras}
\bibliography{literature}

\begin{thebibliography}{}
\makeatletter
\relax
\def\mn@urlcharsother{\let\do\@makeother \do\$\do\&\do\#\do\^\do\_\do\%\do\~}
\def\mn@doi{\begingroup\mn@urlcharsother \@ifnextchar [ {\mn@doi@}
  {\mn@doi@[]}}
\def\mn@doi@[#1]#2{\def\@tempa{#1}\ifx\@tempa\@empty \href
  {http://dx.doi.org/#2} {doi:#2}\else \href {http://dx.doi.org/#2} {#1}\fi
  \endgroup}
\def\mn@eprint#1#2{\mn@eprint@#1:#2::\@nil}
\def\mn@eprint@arXiv#1{\href {http://arxiv.org/abs/#1} {{\tt arXiv:#1}}}
\def\mn@eprint@dblp#1{\href {http://dblp.uni-trier.de/rec/bibtex/#1.xml}
  {dblp:#1}}
\def\mn@eprint@#1:#2:#3:#4\@nil{\def\@tempa {#1}\def\@tempb {#2}\def\@tempc
  {#3}\ifx \@tempc \@empty \let \@tempc \@tempb \let \@tempb \@tempa \fi \ifx
  \@tempb \@empty \def\@tempb {arXiv}\fi \@ifundefined
  {mn@eprint@\@tempb}{\@tempb:\@tempc}{\expandafter \expandafter \csname
  mn@eprint@\@tempb\endcsname \expandafter{\@tempc}}}

\bibitem[\protect\citeauthoryear{Akg{\"u}n, Reisenegger, Mastrano  \&
  Marchant}{Akg{\"u}n et~al.}{2013}]{Akguen2013}
Akg{\"u}n T.,  Reisenegger A.,  Mastrano A.,   Marchant P.,  2013, \mn@doi
  [\mnras] {10.1093/mnras/stt913}, 433, 2445

\bibitem[\protect\citeauthoryear{{Akg{\"u}n}, {Miralles}, {Pons}  \&
  {Cerd{\'a}-Dur{\'a}n}}{{Akg{\"u}n} et~al.}{2016}]{Akgun2016}
{Akg{\"u}n} T.,  {Miralles} J.~A.,  {Pons} J.~A.,   {Cerd{\'a}-Dur{\'a}n} P.,
  2016, \mn@doi [\mnras] {10.1093/mnras/stw1762}, \href
  {http://adsabs.harvard.edu/abs/2016MNRAS.462.1894A} {462, 1894}

\bibitem[\protect\citeauthoryear{{Akg{\"u}n}, {Cerd{\'a}-Dur{\'a}n}, {Miralles}
   \& {Pons}}{{Akg{\"u}n} et~al.}{2017}]{Akgun2017}
{Akg{\"u}n} T.,  {Cerd{\'a}-Dur{\'a}n} P.,  {Miralles} J.~A.,   {Pons} J.~A.,
  2017, \mn@doi [\mnras] {10.1093/mnras/stx2235}, \href
  {http://adsabs.harvard.edu/abs/2017MNRAS.472.3914A} {472, 3914}

\bibitem[\protect\citeauthoryear{{Akg{\"u}n}, {Cerd{\'a}-Dur{\'a}n}, {Miralles}
   \& {Pons}}{{Akg{\"u}n} et~al.}{2018a}]{Akgun2018a}
{Akg{\"u}n} T.,  {Cerd{\'a}-Dur{\'a}n} P.,  {Miralles} J.~A.,   {Pons} J.~A.,
  2018a, \mn@doi [\mnras] {10.1093/mnras/stx2814}, \href
  {http://adsabs.harvard.edu/abs/2018MNRAS.474..625A} {474, 625}

\bibitem[\protect\citeauthoryear{{Akg{\"u}n}, {Cerd{\'a}-Dur{\'a}n}, {Miralles}
   \& {Pons}}{{Akg{\"u}n} et~al.}{2018b}]{Akgun2018b}
{Akg{\"u}n} T.,  {Cerd{\'a}-Dur{\'a}n} P.,  {Miralles} J.~A.,   {Pons} J.~A.,
  2018b, \mn@doi [\mnras] {10.1093/mnras/sty2669}, \href
  {http://adsabs.harvard.edu/abs/2018MNRAS.481.5331A} {481, 5331}

\bibitem[\protect\citeauthoryear{{Alic}, {Moesta}, {Rezzolla}, {Zanotti}  \&
  {Jaramillo}}{{Alic} et~al.}{2012}]{Alic2012}
{Alic} D.,  {Moesta} P.,  {Rezzolla} L.,  {Zanotti} O.,   {Jaramillo} J.~L.,
  2012, \mn@doi [\apj] {10.1088/0004-637X/754/1/36}, 754, 36

\bibitem[\protect\citeauthoryear{{Baiko} \& {Chugunov}}{{Baiko} \&
  {Chugunov}}{2018}]{Baiko2018}
{Baiko} D.~A.,  {Chugunov} A.~I.,  2018, \mn@doi [\mnras]
  {10.1093/mnras/sty2259}, \href
  {https://ui.adsabs.harvard.edu/abs/2018MNRAS.480.5511B} {480, 5511}

\bibitem[\protect\citeauthoryear{{Baumgarte}, {Montero}, {Cordero-Carri{\'o}n}
  \& {M{\"u}ller}}{{Baumgarte} et~al.}{2013}]{Baumgarte2013}
{Baumgarte} T.~W.,  {Montero} P.~J.,  {Cordero-Carri{\'o}n} I.,   {M{\"u}ller}
  E.,  2013, \mn@doi [\prd] {10.1103/PhysRevD.87.044026}, 87, 044026

\bibitem[\protect\citeauthoryear{{Beloborodov}}{{Beloborodov}}{2009}]{Beloborodov2009}
{Beloborodov} A.~M.,  2009, \mn@doi [\apj] {10.1088/0004-637X/703/1/1044},
  \href {http://adsabs.harvard.edu/abs/2009ApJ...703.1044B} {703, 1044}

\bibitem[\protect\citeauthoryear{Beloborodov}{Beloborodov}{2011}]{Beloborodov2011}
Beloborodov A.~M.,  2011, \mn@doi [Astrophysics and Space Science Proceedings]
  {10.1007/978-3-642-17251-9_24}, 21, 299

\bibitem[\protect\citeauthoryear{{Beloborodov}}{{Beloborodov}}{2013a}]{Beloborodov2013b}
{Beloborodov} A.~M.,  2013a, \mn@doi [\apj] {10.1088/0004-637X/762/1/13}, \href
  {https://ui.adsabs.harvard.edu/abs/2013ApJ...762...13B} {762, 13}

\bibitem[\protect\citeauthoryear{{Beloborodov}}{{Beloborodov}}{2013b}]{Beloborodov2013}
{Beloborodov} A.~M.,  2013b, \mn@doi [\apj] {10.1088/0004-637X/777/2/114},
  \href {http://adsabs.harvard.edu/abs/2013ApJ...777..114B} {777, 114}

\bibitem[\protect\citeauthoryear{{Beloborodov} \& {Levin}}{{Beloborodov} \&
  {Levin}}{2014}]{Beloborodov2014}
{Beloborodov} A.~M.,  {Levin} Y.,  2014, \mn@doi [\apj]
  {10.1088/2041-8205/794/2/L24}, \href
  {https://ui.adsabs.harvard.edu/abs/2014ApJ...794L..24B} {794, L24}

\bibitem[\protect\citeauthoryear{{Beloborodov} \& {Thompson}}{{Beloborodov} \&
  {Thompson}}{2007}]{Beloborodov2007}
{Beloborodov} A.~M.,  {Thompson} C.,  2007, \mn@doi [\apj] {10.1086/508917},
  \href {http://adsabs.harvard.edu/abs/2007ApJ...657..967B} {657, 967}

\bibitem[\protect\citeauthoryear{Bernstein, Frieman, Kruskal  \&
  Kulsrud}{Bernstein et~al.}{1958}]{Bernstein1958}
Bernstein I.~B.,  Frieman E.~A.,  Kruskal M.~D.,   Kulsrud R.~M.,  1958,
  \mn@doi [Proceedings of the Royal Society of London Series A]
  {10.1098/rspa.1958.0023}, 244, 17

\bibitem[\protect\citeauthoryear{{Beskin}}{{Beskin}}{2010}]{Beskin2010}
{Beskin} V.~S.,  2010, {MHD Flows in Compact Astrophysical Objects},
  \mn@doi{10.1007/978-3-642-01290-7.
}

\bibitem[\protect\citeauthoryear{{Camenzind}}{{Camenzind}}{2007}]{Camenzind2007}
{Camenzind} M.,  2007, Compact Objects in Astrophysics: White Dwarfs, Neutron
  Stars, and Black Holes.
Springer, \mn@doi{10.1007/978-3-540-49912-1}

\bibitem[\protect\citeauthoryear{{Carrasco} \& {Reula}}{{Carrasco} \&
  {Reula}}{2016}]{Carrasco2016}
{Carrasco} F.~L.,  {Reula} O.~A.,  2016, \mn@doi [\prd]
  {10.1103/PhysRevD.93.085013}, 93, 085013

\bibitem[\protect\citeauthoryear{Carrasco, Vigan{\`o}, Palenzuela  \&
  Pons}{Carrasco et~al.}{2019}]{Carrasco2019}
Carrasco F.,  Vigan{\`o} D.,  Palenzuela C.,   Pons J.~A.,  2019, \mn@doi
  [\mnras] {10.1093/mnrasl/slz016}, 484, L124

\bibitem[\protect\citeauthoryear{{Chugunov} \& {Horowitz}}{{Chugunov} \&
  {Horowitz}}{2010}]{Chugonov2010}
{Chugunov} A.~I.,  {Horowitz} C.~J.,  2010, \mn@doi [\mnras]
  {10.1111/j.1745-3933.2010.00903.x}, \href
  {https://ui.adsabs.harvard.edu/abs/2010MNRAS.407L..54C} {407, L54}

\bibitem[\protect\citeauthoryear{Cline et~al.,}{Cline et~al.}{1980}]{Cline1980}
Cline T.~L.,  et~al., 1980, \mn@doi [\apjl] {10.1086/183221}, 237, L1

\bibitem[\protect\citeauthoryear{Collins, Xu, Norman, Li  \& Li}{Collins
  et~al.}{2010}]{Collins2010}
Collins D.~C.,  Xu H.,  Norman M.~L.,  Li H.,   Li S.,  2010, \mn@doi [\apjs]
  {10.1088/0067-0049/186/2/308}, 186, 308

\bibitem[\protect\citeauthoryear{Coti~Zelati, Rea, Pons, Campana  \&
  Esposito}{Coti~Zelati et~al.}{2018}]{CotiZelati2018}
Coti~Zelati F.,  Rea N.,  Pons J.~A.,  Campana S.,   Esposito P.,  2018,
  \mn@doi [\mnras] {10.1093/mnras/stx2679}, 474, 961

\bibitem[\protect\citeauthoryear{{Dedner}, {Kemm}, {Kr{\"o}ner}, {Munz},
  {Schnitzer}  \& {Wesenberg}}{{Dedner} et~al.}{2002}]{Dedner2002}
{Dedner} A.,  {Kemm} F.,  {Kr{\"o}ner} D.,  {Munz} C.-D.,  {Schnitzer} T.,
  {Wesenberg} M.,  2002, \mn@doi [J. Comput. Phys.] {10.1006/jcph.2001.6961},
  175, 645

\bibitem[\protect\citeauthoryear{{Elenbaas}, {Watts}, {Turolla}  \&
  {Heyl}}{{Elenbaas} et~al.}{2016}]{Elenbaas2016}
{Elenbaas} C.,  {Watts} A.~L.,  {Turolla} R.,   {Heyl} J.~S.,  2016, \mn@doi
  [\mnras] {10.1093/mnras/stv2860}, \href
  {https://ui.adsabs.harvard.edu/abs/2016MNRAS.456.3282E} {456, 3282}

\bibitem[\protect\citeauthoryear{Etienne, Wan, Babiuc, McWilliams  \&
  Choudhary}{Etienne et~al.}{2017}]{Etienne2017}
Etienne Z.~B.,  Wan M.-B.,  Babiuc M.~C.,  McWilliams S.~T.,   Choudhary A.,
  2017, \mn@doi [Classical and Quantum Gravity] {10.1088/1361-6382/aa8ab3}, 34,
  215001

\bibitem[\protect\citeauthoryear{Fern{\'a}ndez \& Davis}{Fern{\'a}ndez \&
  Davis}{2011}]{Fernandez2011}
Fern{\'a}ndez R.,  Davis S.~W.,  2011, \mn@doi [\apj]
  {10.1088/0004-637X/730/2/131}, 730, 131

\bibitem[\protect\citeauthoryear{{Fujisawa} \& {Kisaka}}{{Fujisawa} \&
  {Kisaka}}{2014}]{Fujisawa2014}
{Fujisawa} K.,  {Kisaka} S.,  2014, \mn@doi [\mnras] {10.1093/mnras/stu1911},
  \href {http://adsabs.harvard.edu/abs/2014MNRAS.445.2777F} {445, 2777}

\bibitem[\protect\citeauthoryear{{Gabler}, {Cerd{\'a}-Dur{\'a}n},
  {Stergioulas}, {Font}  \& {M{\"u}ller}}{{Gabler} et~al.}{2012}]{Gabler2012}
{Gabler} M.,  {Cerd{\'a}-Dur{\'a}n} P.,  {Stergioulas} N.,  {Font} J.~A.,
  {M{\"u}ller} E.,  2012, \mn@doi [\mnras] {10.1111/j.1365-2966.2012.20454.x},
  \href {https://ui.adsabs.harvard.edu/abs/2012MNRAS.421.2054G} {421, 2054}

\bibitem[\protect\citeauthoryear{{Gill} \& {Heyl}}{{Gill} \&
  {Heyl}}{2010}]{Gill2010}
{Gill} R.,  {Heyl} J.~S.,  2010, \mn@doi [\mnras]
  {10.1111/j.1365-2966.2010.17038.x}, \href
  {https://ui.adsabs.harvard.edu/abs/2010MNRAS.407.1926G} {407, 1926}

\bibitem[\protect\citeauthoryear{{Glampedakis}, {Lander}  \&
  {Andersson}}{{Glampedakis} et~al.}{2014}]{Glampedakis2014}
{Glampedakis} K.,  {Lander} S.~K.,   {Andersson} N.,  2014, \mn@doi [\mnras]
  {10.1093/mnras/stt1814}, \href
  {http://adsabs.harvard.edu/abs/2014MNRAS.437....2G} {437, 2}

\bibitem[\protect\citeauthoryear{{Goldreich} \& {Julian}}{{Goldreich} \&
  {Julian}}{1969}]{GJ1969}
{Goldreich} P.,  {Julian} W.~H.,  1969, \mn@doi [\apj] {10.1086/150119}, \href
  {http://adsabs.harvard.edu/abs/1969ApJ...157..869G} {157, 869}

\bibitem[\protect\citeauthoryear{{Goldreich} \& {Reisenegger}}{{Goldreich} \&
  {Reisenegger}}{1992}]{Goldreich1992}
{Goldreich} P.,  {Reisenegger} A.,  1992, \mn@doi [\apj] {10.1086/171646},
  \href {https://ui.adsabs.harvard.edu/abs/1992ApJ...395..250G} {395, 250}

\bibitem[\protect\citeauthoryear{Goodale, Allen, Lanfermann, Mass{\'o}, Radke,
  Seidel  \& Shalf}{Goodale et~al.}{2003}]{Goodale2002a}
Goodale T.,  Allen G.,  Lanfermann G.,  Mass{\'o} J.,  Radke T.,  Seidel E.,
  Shalf J.,  2003, in Vector and Parallel Processing -- VECPAR'2002, 5th
  International Conference, Lecture Notes in Computer Science. Springer,
  Berlin, \url {http://edoc.mpg.de/3341}

\bibitem[\protect\citeauthoryear{{Goodman}}{{Goodman}}{1986}]{Goodman_1986ApJ...308L..47}
{Goodman} J.,  1986, \mn@doi [\apjl] {10.1086/184741}, \href
  {https://ui.adsabs.harvard.edu/abs/1986ApJ...308L..47G} {308, L47}

\bibitem[\protect\citeauthoryear{{Gourgouliatos}, {Wood}  \&
  {Hollerbach}}{{Gourgouliatos} et~al.}{2016}]{Gourgouliatos2016}
{Gourgouliatos} K.~N.,  {Wood} T.~S.,   {Hollerbach} R.,  2016, \mn@doi
  [Proceedings of the National Academy of Science] {10.1073/pnas.1522363113},
  \href {https://ui.adsabs.harvard.edu/abs/2016PNAS..113.3944G} {113, 3944}

\bibitem[\protect\citeauthoryear{Grad \& Rubin}{Grad \& Rubin}{1958}]{Grad1958}
Grad H.,  Rubin H.,  1958, Proceedings of the Second United Nations Conference
  on the Peaceful Uses of Atomic Energy (Geneva), 31, 190

\bibitem[\protect\citeauthoryear{{Hasco{\"e}t}, {Beloborodov}  \& {den
  Hartog}}{{Hasco{\"e}t} et~al.}{2014}]{Hascoet2014}
{Hasco{\"e}t} R.,  {Beloborodov} A.~M.,   {den Hartog} P.~R.,  2014, \mn@doi
  [\apj] {10.1088/2041-8205/786/1/L1}, \href
  {https://ui.adsabs.harvard.edu/abs/2014ApJ...786L...1H} {786, L1}

\bibitem[\protect\citeauthoryear{{Horowitz} \& {Kadau}}{{Horowitz} \&
  {Kadau}}{2009}]{Horowitz2009}
{Horowitz} C.~J.,  {Kadau} K.,  2009, \mn@doi [\prl]
  {10.1103/PhysRevLett.102.191102}, \href
  {https://ui.adsabs.harvard.edu/\#abs/2009PhRvL.102s1102H} {102, 191102}

\bibitem[\protect\citeauthoryear{Hurley et~al.,}{Hurley
  et~al.}{1999}]{Hurley1999}
Hurley K.,  et~al., 1999, \mn@doi [\nat] {10.1038/16199}, 397, 41

\bibitem[\protect\citeauthoryear{Hurley et~al.,}{Hurley
  et~al.}{2005}]{Hurley2005}
Hurley K.,  et~al., 2005, \mn@doi [\nat] {10.1038/nature03519}, 434, 1098

\bibitem[\protect\citeauthoryear{{Janka}, {Aloy}, {Mazzali}  \& {Pian}}{{Janka}
  et~al.}{2006}]{Janka_2006ApJ...645.1305}
{Janka} H.~T.,  {Aloy} M.~A.,  {Mazzali} P.~A.,   {Pian} E.,  2006, \mn@doi
  [\apj] {10.1086/504580}, \href
  {https://ui.adsabs.harvard.edu/abs/2006ApJ...645.1305J} {645, 1305}

\bibitem[\protect\citeauthoryear{{Jones}}{{Jones}}{1988}]{Jones1988}
{Jones} P.~B.,  1988, \mn@doi [\mnras] {10.1093/mnras/233.4.875}, \href
  {https://ui.adsabs.harvard.edu/abs/1988MNRAS.233..875J} {233, 875}

\bibitem[\protect\citeauthoryear{Kaspi \& Beloborodov}{Kaspi \&
  Beloborodov}{2017}]{Kaspi2017}
Kaspi V.~M.,  Beloborodov A.~M.,  2017, \mn@doi [\araa]
  {10.1146/annurev-astro-081915-023329}, 55, 261

\bibitem[\protect\citeauthoryear{{Kojima}}{{Kojima}}{2017}]{Kojima2017}
{Kojima} Y.,  2017, \mn@doi [\mnras] {10.1093/mnras/stx584}, \href
  {http://adsabs.harvard.edu/abs/2017MNRAS.468.2011K} {468, 2011}

\bibitem[\protect\citeauthoryear{{Kojima}}{{Kojima}}{2018}]{Kojima2018}
{Kojima} Y.,  2018, \mn@doi [\mnras] {10.1093/mnras/sty866}, \href
  {http://adsabs.harvard.edu/abs/2018MNRAS.477.3530K} {477, 3530}

\bibitem[\protect\citeauthoryear{{Kojima} \& {Okamoto}}{{Kojima} \&
  {Okamoto}}{2018}]{Kojima2018b}
{Kojima} Y.,  {Okamoto} S.,  2018, \mn@doi [\mnras] {10.1093/mnras/sty176},
  \href {http://adsabs.harvard.edu/abs/2018MNRAS.475.5290K} {475, 5290}

\bibitem[\protect\citeauthoryear{Komissarov}{Komissarov}{2004}]{Komissarov2004}
Komissarov S.~S.,  2004, \mn@doi [\mnras] {10.1111/j.1365-2966.2004.07598.x},
  350, 427–448

\bibitem[\protect\citeauthoryear{{Komissarov}}{{Komissarov}}{2011}]{Komissarov2011}
{Komissarov} S.~S.,  2011, \mn@doi [\mnras] {10.1111/j.1745-3933.2011.01150.x},
  418, L94

\bibitem[\protect\citeauthoryear{{Landau} \& {Lifshitz}}{{Landau} \&
  {Lifshitz}}{2012}]{Landau:elasticity}
{Landau} L.~D.,  {Lifshitz} E.~M.,  2012, {Theory of elasticity}.
Elsevier

\bibitem[\protect\citeauthoryear{{Lander} \& {Gourgouliatos}}{{Lander} \&
  {Gourgouliatos}}{2019}]{Lander2019}
{Lander} S.~K.,  {Gourgouliatos} K.~N.,  2019, \mn@doi [\mnras]
  {10.1093/mnras/stz1042}, \href
  {https://ui.adsabs.harvard.edu/abs/2019MNRAS.486.4130L} {486, 4130}

\bibitem[\protect\citeauthoryear{{Lander}, {Andersson}, {Antonopoulou}  \&
  {Watts}}{{Lander} et~al.}{2015}]{Lander2015}
{Lander} S.~K.,  {Andersson} N.,  {Antonopoulou} D.,   {Watts} A.~L.,  2015,
  \mn@doi [\mnras] {10.1093/mnras/stv432}, \href
  {https://ui.adsabs.harvard.edu/abs/2015MNRAS.449.2047L} {449, 2047}

\bibitem[\protect\citeauthoryear{{Levin} \& {Lyutikov}}{{Levin} \&
  {Lyutikov}}{2012}]{Levin2012}
{Levin} Y.,  {Lyutikov} M.,  2012, \mn@doi [\mnras]
  {10.1111/j.1365-2966.2012.22016.x}, \href
  {https://ui.adsabs.harvard.edu/abs/2012MNRAS.427.1574L} {427, 1574}

\bibitem[\protect\citeauthoryear{Li \& Beloborodov}{Li \&
  Beloborodov}{2015}]{Li2015}
Li X.,  Beloborodov A.~M.,  2015, \mn@doi [\apj] {10.1088/0004-637X/815/1/25},
  815, 25

\bibitem[\protect\citeauthoryear{{Li}, {Levin}  \& {Beloborodov}}{{Li}
  et~al.}{2016}]{Li2016}
{Li} X.,  {Levin} Y.,   {Beloborodov} A.~M.,  2016, \mn@doi [\apj]
  {10.3847/1538-4357/833/2/189}, \href
  {https://ui.adsabs.harvard.edu/abs/2016ApJ...833..189L} {833, 189}

\bibitem[\protect\citeauthoryear{Li, Zrake  \& Beloborodov}{Li
  et~al.}{2018}]{Li2018}
Li X.,  Zrake J.,   Beloborodov A.~M.,  2018, arXiv e-prints

\bibitem[\protect\citeauthoryear{{Link}}{{Link}}{2014}]{Link2014}
{Link} B.,  2014, \mn@doi [\mnras] {10.1093/mnras/stu584}, \href
  {https://ui.adsabs.harvard.edu/abs/2014MNRAS.441.2676L} {441, 2676}

\bibitem[\protect\citeauthoryear{{L{\"o}ffler} et~al.,}{{L{\"o}ffler}
  et~al.}{2012}]{Loeffler2012}
{L{\"o}ffler} F.,  et~al., 2012, \mn@doi [Classical and Quantum Gravity]
  {10.1088/0264-9381/29/11/115001}, 29, 115001

\bibitem[\protect\citeauthoryear{{L{\"u}st} \& {Schl{\"u}ter}}{{L{\"u}st} \&
  {Schl{\"u}ter}}{1954}]{Luest1954}
{L{\"u}st} R.,  {Schl{\"u}ter} A.,  1954, \zap, \href
  {http://adsabs.harvard.edu/abs/1954ZA.....34..263L} {34, 263}

\bibitem[\protect\citeauthoryear{Lyutikov}{Lyutikov}{2003}]{Lyutikov2003}
Lyutikov M.,  2003, \mn@doi [\mnras] {10.1046/j.1365-2966.2003.07110.x}, 346,
  540

\bibitem[\protect\citeauthoryear{{Lyutikov}}{{Lyutikov}}{2015}]{Lyutikov2015}
{Lyutikov} M.,  2015, \mn@doi [\mnras] {10.1093/mnras/stu2413}, \href
  {https://ui.adsabs.harvard.edu/abs/2015MNRAS.447.1407L} {447, 1407}

\bibitem[\protect\citeauthoryear{{McKinney}}{{McKinney}}{2006}]{McKinney2006}
{McKinney} J.~C.,  2006, \mn@doi [\mnras] {10.1111/j.1365-2966.2006.10087.x},
  367, 1797

\bibitem[\protect\citeauthoryear{Mereghetti, Pons  \& Melatos}{Mereghetti
  et~al.}{2015}]{Mereghetti2015}
Mereghetti S.,  Pons J.~A.,   Melatos A.,  2015, \mn@doi [\ssr]
  {10.1007/s11214-015-0146-y}, 191, 315

\bibitem[\protect\citeauthoryear{{M{\'e}sz{\'a}ros} \&
  {Rees}}{{M{\'e}sz{\'a}ros} \& {Rees}}{2000}]{Meszaros:2000ApJ...530..292}
{M{\'e}sz{\'a}ros} P.,  {Rees} M.~J.,  2000, \mn@doi [\apj] {10.1086/308371},
  530, 292

\bibitem[\protect\citeauthoryear{{Meszaros}, {Laguna}  \& {Rees}}{{Meszaros}
  et~al.}{1993}]{Meszaros_1993ApJ...415..181}
{Meszaros} P.,  {Laguna} P.,   {Rees} M.~J.,  1993, \mn@doi [\apj]
  {10.1086/173154}, \href
  {https://ui.adsabs.harvard.edu/abs/1993ApJ...415..181M} {415, 181}

\bibitem[\protect\citeauthoryear{{Mignone} \& {Tzeferacos}}{{Mignone} \&
  {Tzeferacos}}{2010}]{Mignone2010}
{Mignone} A.,  {Tzeferacos} P.,  2010, \mn@doi [\jcop]
  {10.1016/j.jcp.2009.11.026}, 229, 2117

\bibitem[\protect\citeauthoryear{{Mikic} \& {Linker}}{{Mikic} \&
  {Linker}}{1994}]{Mikic1994}
{Mikic} Z.,  {Linker} J.~A.,  1994, \mn@doi [\apj] {10.1086/174460}, \href
  {https://ui.adsabs.harvard.edu/abs/1994ApJ...430..898M} {430, 898}

\bibitem[\protect\citeauthoryear{{Miranda-Aranguren}, {Aloy}  \&
  {Rembiasz}}{{Miranda-Aranguren} et~al.}{2018}]{Miranda-Aranguren2018}
{Miranda-Aranguren} S.,  {Aloy} M.~A.,   {Rembiasz} T.,  2018, \mn@doi [\mnras]
  {10.1093/mnras/sty419}, \href
  {https://ui.adsabs.harvard.edu/abs/2018MNRAS.476.3837M} {476, 3837}

\bibitem[\protect\citeauthoryear{Montero, Baumgarte  \& M{\"u}ller}{Montero
  et~al.}{2014}]{Montero2014}
Montero P.~J.,  Baumgarte T.~W.,   M{\"u}ller E.,  2014, \mn@doi [\prd]
  {10.1103/PhysRevD.89.084043}, 89, 084043

\bibitem[\protect\citeauthoryear{{Nakar}, {Piran}  \& {Sari}}{{Nakar}
  et~al.}{2005}]{Nakar_2005ApJ...635..516}
{Nakar} E.,  {Piran} T.,   {Sari} R.,  2005, \mn@doi [\apj] {10.1086/497296},
  \href {https://ui.adsabs.harvard.edu/abs/2005ApJ...635..516N} {635, 516}

\bibitem[\protect\citeauthoryear{{Paczynski}}{{Paczynski}}{1986}]{Paczynski_1986ApJ...308L..43}
{Paczynski} B.,  1986, \mn@doi [\apjl] {10.1086/184740}, \href
  {https://ui.adsabs.harvard.edu/abs/1986ApJ...308L..43P} {308, L43}

\bibitem[\protect\citeauthoryear{{Palenzuela}, {Lehner}, {Reula}  \&
  {Rezzolla}}{{Palenzuela} et~al.}{2009}]{Palenzuela2009}
{Palenzuela} C.,  {Lehner} L.,  {Reula} O.,   {Rezzolla} L.,  2009, \mn@doi
  [\mnras] {10.1111/j.1365-2966.2009.14454.x}, 394, 1727

\bibitem[\protect\citeauthoryear{{Palenzuela}, {Garrett}, {Lehner}  \&
  {Liebling}}{{Palenzuela} et~al.}{2010}]{Palenzuela2010}
{Palenzuela} C.,  {Garrett} T.,  {Lehner} L.,   {Liebling} S.~L.,  2010,
  \mn@doi [\prd] {10.1103/PhysRevD.82.044045}, 82, 044045

\bibitem[\protect\citeauthoryear{Parfrey, Beloborodov  \& Hui}{Parfrey
  et~al.}{2012}]{Parfrey2012}
Parfrey K.,  Beloborodov A.~M.,   Hui L.,  2012, \mn@doi [\apjl]
  {10.1088/2041-8205/754/1/L12}, 754, L12

\bibitem[\protect\citeauthoryear{Parfrey, Beloborodov  \& Hui}{Parfrey
  et~al.}{2013}]{Parfrey2013}
Parfrey K.,  Beloborodov A.~M.,   Hui L.,  2013, \mn@doi [\apj]
  {10.1088/0004-637X/774/2/92}, 774, 92

\bibitem[\protect\citeauthoryear{{Parfrey}, {Spitkovsky}  \&
  {Beloborodov}}{{Parfrey} et~al.}{2017}]{Parfrey2017}
{Parfrey} K.,  {Spitkovsky} A.,   {Beloborodov} A.~M.,  2017, \mn@doi [\mnras]
  {10.1093/mnras/stx950}, 469, 3656

\bibitem[\protect\citeauthoryear{{Paschalidis} \& {Shapiro}}{{Paschalidis} \&
  {Shapiro}}{2013}]{Paschalidis2013}
{Paschalidis} V.,  {Shapiro} S.~L.,  2013, \mn@doi [\prd]
  {10.1103/PhysRevD.88.104031}, \href
  {http://adsabs.harvard.edu/abs/2013PhRvD..88j4031P} {88, 104031}

\bibitem[\protect\citeauthoryear{{Perna} \& {Pons}}{{Perna} \&
  {Pons}}{2011}]{Perna2011}
{Perna} R.,  {Pons} J.~A.,  2011, \mn@doi [\apj] {10.1088/2041-8205/727/2/L51},
  \href {https://ui.adsabs.harvard.edu/abs/2011ApJ...727L..51P} {727, L51}

\bibitem[\protect\citeauthoryear{{Pili}, {Bucciantini}  \& {Del Zanna}}{{Pili}
  et~al.}{2015}]{Pili2015}
{Pili} A.~G.,  {Bucciantini} N.,   {Del Zanna} L.,  2015, \mn@doi [\mnras]
  {10.1093/mnras/stu2628}, \href
  {http://adsabs.harvard.edu/abs/2015MNRAS.447.2821P} {447, 2821}

\bibitem[\protect\citeauthoryear{{Piran}, {Shemi}  \& {Narayan}}{{Piran}
  et~al.}{1993}]{Piran_1993MNRAS.263..861}
{Piran} T.,  {Shemi} A.,   {Narayan} R.,  1993, \mn@doi [\mnras]
  {10.1093/mnras/263.4.861}, \href
  {https://ui.adsabs.harvard.edu/abs/1993MNRAS.263..861P} {263, 861}

\bibitem[\protect\citeauthoryear{Pons \& Geppert}{Pons \&
  Geppert}{2007}]{Pons2007}
Pons J.~A.,  Geppert U.,  2007, \mn@doi [\aap] {10.1051/0004-6361:20077456},
  470, 303

\bibitem[\protect\citeauthoryear{{Pons}, {Miralles}  \& {Geppert}}{{Pons}
  et~al.}{2009}]{Pons2009}
{Pons} J.~A.,  {Miralles} J.~A.,   {Geppert} U.,  2009, \mn@doi [\aap]
  {10.1051/0004-6361:200811229}, \href
  {https://ui.adsabs.harvard.edu/abs/2009A&A...496..207P} {496, 207}

\bibitem[\protect\citeauthoryear{Rea \& Esposito}{Rea \&
  Esposito}{2011}]{Rea2011}
Rea N.,  Esposito P.,  2011, \mn@doi [Astrophysics and Space Science
  Proceedings] {10.1007/978-3-642-17251-9_21}, 21, 247

\bibitem[\protect\citeauthoryear{Roumeliotis, Sturrock  \&
  Antiochos}{Roumeliotis et~al.}{1994}]{Roumeliotis1994}
Roumeliotis G.,  Sturrock P.~A.,   Antiochos S.~K.,  1994, \mn@doi [\apj]
  {10.1086/173862}, 423, 847

\bibitem[\protect\citeauthoryear{Shafranov}{Shafranov}{1966}]{Shafranov1966}
Shafranov V.,  1966, Reviews of Plasma Physics, 2, 103

\bibitem[\protect\citeauthoryear{{Shemi} \& {Piran}}{{Shemi} \&
  {Piran}}{1990}]{Shemi_1990ApJ...365L..55}
{Shemi} A.,  {Piran} T.,  1990, \mn@doi [\apjl] {10.1086/185887}, \href
  {https://ui.adsabs.harvard.edu/abs/1990ApJ...365L..55S} {365, L55}

\bibitem[\protect\citeauthoryear{Shibata}{Shibata}{2015}]{Shibata2015}
Shibata M.,  2015, Numerical Relativity.
100 Years of General Relativity, World Scientific Publishing Company

\bibitem[\protect\citeauthoryear{Spitkovsky}{Spitkovsky}{2006}]{Spitkovsky2006}
Spitkovsky A.,  2006, \mn@doi [\apjl] {10.1086/507518}, 648, L51

\bibitem[\protect\citeauthoryear{{Steiner} \& {Watts}}{{Steiner} \&
  {Watts}}{2009}]{Steiner2009}
{Steiner} A.~W.,  {Watts} A.~L.,  2009, \mn@doi [Physical Review Letters]
  {10.1103/PhysRevLett.103.181101}, \href
  {http://adsabs.harvard.edu/abs/2009PhRvL.103r1101S} {103, 181101}

\bibitem[\protect\citeauthoryear{{Suresh} \& {Huynh}}{{Suresh} \&
  {Huynh}}{1997}]{Suresh1997}
{Suresh} A.,  {Huynh} H.~T.,  1997, \mn@doi [Journal of Computational Physics]
  {10.1006/jcph.1997.5745}, 136, 83

\bibitem[\protect\citeauthoryear{Tayler}{Tayler}{1973}]{Tayler1973}
Tayler R.~J.,  1973, \mn@doi [\mnras] {10.1093/mnras/161.4.365}, 161, 365

\bibitem[\protect\citeauthoryear{Thompson \& Duncan}{Thompson \&
  Duncan}{1995}]{Thompson1995}
Thompson C.,  Duncan R.~C.,  1995, \mn@doi [\mnras] {10.1093/mnras/275.2.255},
  275, 255

\bibitem[\protect\citeauthoryear{Thompson \& Duncan}{Thompson \&
  Duncan}{1996}]{Thompson1996}
Thompson C.,  Duncan R.~C.,  1996, \mn@doi [\apj] {10.1086/178147}, 473, 322

\bibitem[\protect\citeauthoryear{Thompson \& Duncan}{Thompson \&
  Duncan}{2001}]{Thompson2001}
Thompson C.,  Duncan R.~C.,  2001, \mn@doi [\apj] {10.1086/323256}, 561, 980

\bibitem[\protect\citeauthoryear{{Thompson}, {Lyutikov}  \&
  {Kulkarni}}{{Thompson} et~al.}{2002}]{Thompson2002}
{Thompson} C.,  {Lyutikov} M.,   {Kulkarni} S.~R.,  2002, \mn@doi [\apj]
  {10.1086/340586}, \href
  {https://ui.adsabs.harvard.edu/abs/2002ApJ...574..332T} {574, 332}

\bibitem[\protect\citeauthoryear{{Thompson}, {Yang}  \& {Ortiz}}{{Thompson}
  et~al.}{2017}]{Thompson2017}
{Thompson} C.,  {Yang} H.,   {Ortiz} N.,  2017, \mn@doi [\apj]
  {10.3847/1538-4357/aa6c30}, \href
  {https://ui.adsabs.harvard.edu/abs/2017ApJ...841...54T} {841, 54}

\bibitem[\protect\citeauthoryear{Turolla, Zane  \& Watts}{Turolla
  et~al.}{2015}]{Turolla2015}
Turolla R.,  Zane S.,   Watts A.~L.,  2015, \mn@doi [Reports on Progress in
  Physics] {10.1088/0034-4885/78/11/116901}, 78, 116901

\bibitem[\protect\citeauthoryear{Uchida}{Uchida}{1997}]{Uchida1997}
Uchida T.,  1997, \mn@doi [Phys. Rev. E] {10.1103/physreve.56.2181}, 56,
  2181–2197

\bibitem[\protect\citeauthoryear{Vigan{\`o}, Pons  \& Miralles}{Vigan{\`o}
  et~al.}{2011}]{Vigano2011}
Vigan{\`o} D.,  Pons J.~A.,   Miralles J.~A.,  2011, \mn@doi [\aap]
  {10.1051/0004-6361/201117105}, 533, A125

\bibitem[\protect\citeauthoryear{Vigan{\`o}, Pons  \& Miralles}{Vigan{\`o}
  et~al.}{2012}]{Vigano2012}
Vigan{\`o} D.,  Pons J.~A.,   Miralles J.~A.,  2012, \mn@doi [Computer Physics
  Communications] {10.1016/j.cpc.2012.04.029}, 183, 2042

\bibitem[\protect\citeauthoryear{{Vigan{\`o}}, {Rea}, {Pons}, {Perna},
  {Aguilera}  \& {Miralles}}{{Vigan{\`o}} et~al.}{2013}]{Vigano2013}
{Vigan{\`o}} D.,  {Rea} N.,  {Pons} J.~A.,  {Perna} R.,  {Aguilera} D.~N.,
  {Miralles} J.~A.,  2013, \mn@doi [\mnras] {10.1093/mnras/stt1008}, \href
  {https://ui.adsabs.harvard.edu/abs/2013MNRAS.434..123V} {434, 123}

\bibitem[\protect\citeauthoryear{Wald}{Wald}{2010}]{Wald2010}
Wald R.,  2010, General Relativity.
University of Chicago Press

\bibitem[\protect\citeauthoryear{Woods \& Thompson}{Woods \&
  Thompson}{2006}]{Woods2006}
Woods P.~M.,  Thompson C.,  2006, Soft gamma repeaters and anomalous X-ray
  pulsars: magnetar candidates.
pp 547--586

\makeatother
\end{thebibliography}

\appendix

\section{Numerical details}
\label{sec:num_details} 

\subsection{The augmented system}
\label{sec:augmented_system} 

%
\begin{figure}
	\centering
	\includegraphics[width=0.49\textwidth]{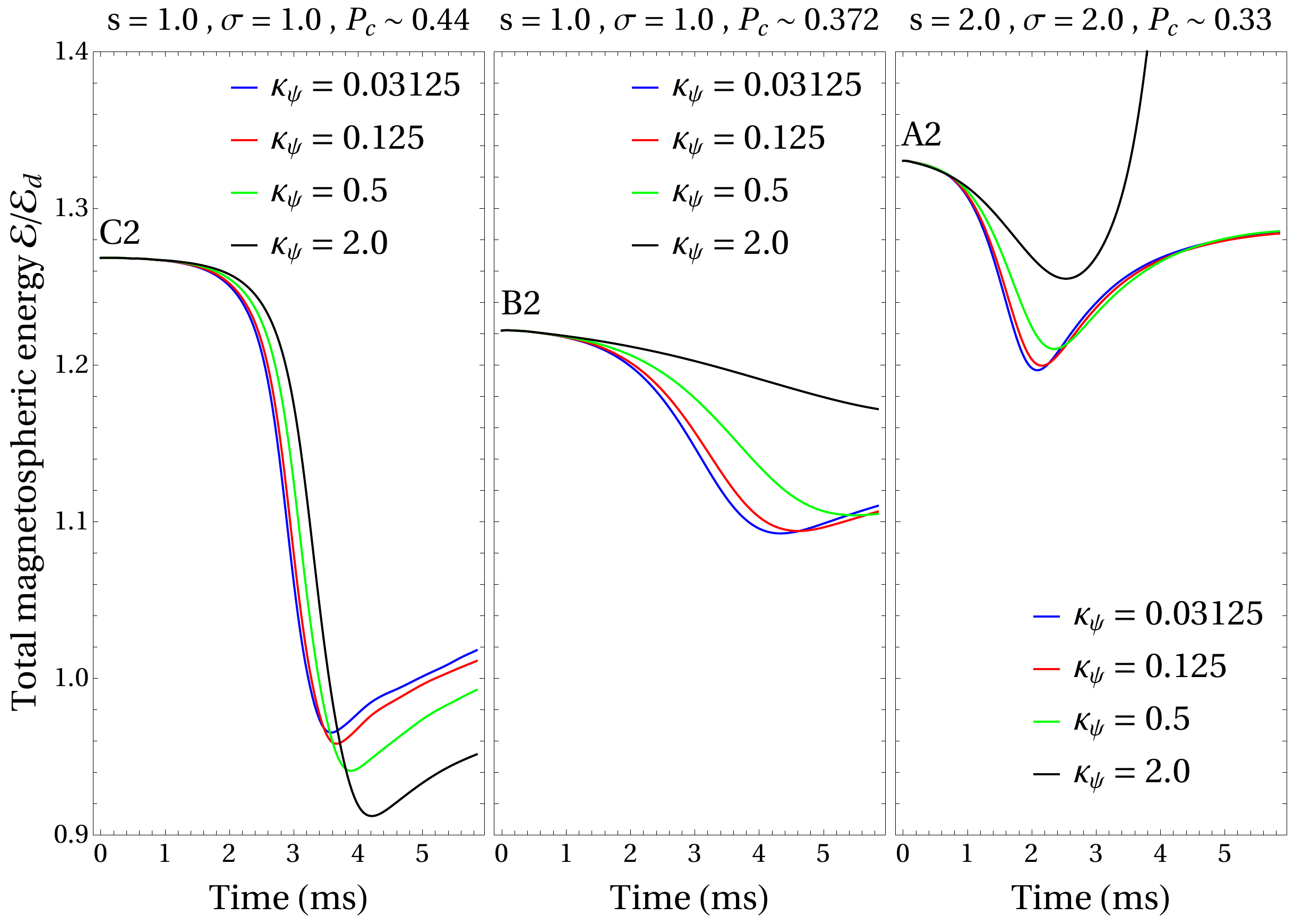} 
	\vspace{-11pt}
	\caption{Energy evolution of the high energy initial data models A2, B2, and C2 using different damping constants $\kappa_\psi$ (divergence cleaning) in a low resolution study (16 points per $R_*$). While one observes a converging evolution for the lower cleaning potentials $\kappa_\psi=0.03125$ and $\kappa_\psi=0.125$, the energy evolution shows a strong (non-physical) dependence on $\kappa_\psi$ for larger damping constants. This effect is amplified in the high resolution (32 points per $R_*$).}
	\label{fig:psi_optimization}
\end{figure}

In order to preserve the physical conditions $\text{div}\vect{\BB}=0$ and $\text{div}\vect{\EE}=\rrho$ we make use of hyperbolic/parabolic cleaning potentials \citep{Dedner2002,Palenzuela2009,Mignone2010}. Specifically, we implement an augmented system of Maxwell's equations as follows \citep{Palenzuela2009,Miranda-Aranguren2018}:
\begin{align}
\partial_t\phi-\partial_i \EE^i&=-\tilde{\rho}_{\rm e}-\kappa_\phi\phi\label{eq:scalar_phi}\\
\partial_t \EE^i-\partial_j\left(\epsilon^{ijk}\BB_k+\delta^{ij}\phi\right)&=-\JJ^i_{\rm \textsc{FF}}\label{eq:vector_phi}\\
\partial_t\psi+\partial_i \BB^i&=-\kappa_\psi\psi\label{eq:scalar_psi}\\
\partial_t \BB^i+\partial_j\left(\epsilon^{ijk}\EE_k+\delta^{ij}\psi\right)&=0\label{eq:vector_psi}
\end{align}
Here, $\psi$ (divergence cleaning) and $\phi$ (charge conservation) are the scalar potentials, $\kappa_\phi$ and $\kappa_\psi$ the respective damping constants and $\delta^{ij}$ denotes the Kronecker delta. As for the practical implementation, we follow a Strang 
splitting approach \citep[as employed, e.g. in][]{Komissarov2004}, effectively solving part of the scalar equations (\ref{eq:scalar_phi}) and (\ref{eq:scalar_psi}) analytically. Prior (before \texttt{MoL\_Step}) and after (before \texttt{MoL\_PostStep}) the time integration of the Einstein Toolkit \texttt{thorn} \texttt{MoL} we evolve in time the equations
\begin{align}
\phi\left(t\right)&=\phi_0\exp\left[-\kappa_\phi t\right],\\
\psi\left(t\right)&=\psi_0\exp\left[-\kappa_\psi t\right],
\end{align}
for a time $t=\Delta t/2$. The coefficients $\kappa_\phi$ and $\kappa_\psi$ have to be chosen by optimisation in accordance with the grid properties. 

We find it beneficial to choose a large value for $\kappa_\phi$, effectively dissipating charge conservation errors on very short timescales. As for the divergence cleaning, we conducted a series of tests, optimizing $\kappa_\psi$ to yield stable and converging evolution for all shown resolutions, ultimately resorting to $\kappa_\psi = 0.125$ (see Figure~\ref{fig:psi_optimization} for a review of the optimisation process).

It should be noted at this point that \citet{Mignone2010} present a promising scheme of choosing $\kappa_\psi$ according to the grid resolution that has also been used in \cite{Miranda-Aranguren2018}. In the framework of mesh refinement of the Einstein Toolkit, this would result in a different damping of the cleaning potentials across the refinement levels. We have found that the optimisation of the hyperbolic/parabolic cleaning becomes a very subtle issue and may experience strong numerical effects when increasing the overall resolution. This observation may, however, be an artifact of the fixed boundary of the magnetar surface - which on a Cartesian grid, resembles an accumulation of boxes rather than a perfectly aligned spherical boundary. The exploration of these effects and the transition to a fully spherical version of this force-free \texttt{thorn} \citep[as introduced in][]{Baumgarte2013,Montero2014} will be a subject of future efforts.

\subsection{Conservation of force-free constraints}
\label{sec:ff_breakdown} 

FFE codes are valid in the limit of high electromagnetic energy compared to the rest mass and thermal energy of the respective plasma. The dynamics of force-free fields is described entirely without the plasma four-velocity. However, demanding the existence of a physical, timelike velocity field $\vect{u}$ with $F_{\mu\nu}u^\nu = 0$, as well as the degeneracy condition $F_{\mu\nu}J^\nu=0$ \citep[see][for a detailed algebraic review]{Uchida1997} one is left with the aforementioned constraints:
\begin{align}
\vect{\EE}\cdot\vect{\BB}&=0 \tag{\ref{eq:force_free_crossfield}}\\
\vect{\BB}^2-\vect{\EE}^2&\geq 0 \tag{\ref{eq:force_free_dominance}}
\end{align}
Within the shown simulations we find it beneficial to employ an approach presented in \citet{Komissarov2011} and \citet{Parfrey2017} in order to archive $\partial_t\left(\vect{\EE}\cdot\vect{\BB}\right)=0$ throughout the evolution (by making use of the force-free current as in equation~\ref{eq:ff_current}) without the employment of target currents \citep[as discussed in][]{Parfrey2017}. Additionally, we include a suitable Ohm's law \citep[][section C3]{Komissarov2004} into our Strang splitting approach aiming towards an evolution minimizing the violation of conditions (\ref{eq:force_free_crossfield}), and (\ref{eq:force_free_dominance}). 

In order to build up a force-free current, \citet{Komissarov2004} introduces a generalised Ohm's law in the context of FFE:
\begin{align}
\begin{split}
\vect{\JJ}=\sigma_\parallel\vect{\EE}_\parallel+\sigma_\perp\vect{\EE}_\perp+\vect{\tilde{j}}_d\label{eq:GeneralizedOhmsLaw},
\end{split}
\end{align} 
where the subscripts $\parallel$ and $\perp$ denote the components parallel and perpendicular to the magnetic field, $\vect{\BB}$. A to be specified model for $\sigma$ introduces a suitable resistivity into the force-free system \citep[see also][for further comments on resistive FFE]{Lyutikov2003}, while $\vect{\tilde{j}}_d$ is the drift current perpendicular to the electric and magnetic fields. In its general form, (\ref{eq:GeneralizedOhmsLaw}) plays the central role in ensuring the force-free conditions (\ref{eq:force_free_crossfield}) and (\ref{eq:force_free_dominance}). \citet{Komissarov2004} suggests a resistivity model that depends on the time-step of the evolution $\Delta t$ (throughout the presented simulations we employ CFL $=0.2$), where
\begin{align}
\begin{split}
\sigma_\parallel = \frac{d}{\Delta t}\label{eq:sigmapar}.
\end{split}
\end{align}
The cross-field resistivity $\sigma_\perp$ is strongly linked to the violation of condition (\ref{eq:force_free_dominance}),
\begin{align}
\arraycolsep=1.4pt\def\arraystretch{1.5}
\begin{split}
\sigma_\perp=\left\{\begin{array}{ccc}
0 & \qquad :\qquad & \vect{B}^2\geq\vect{E}^2\\
b\displaystyle \frac{\left(\EE_\perp-\EE_\perp^*\right)}{\EE_\perp^*} & \qquad :\qquad & \vect{\BB}^2<\vect{\EE}^2
\end{array}\right.\label{eq:sigmaperp},
\end{split}
\end{align}
where $\EE_\perp=\left|\vect{\EE}_\perp\right|$ and $\left(\EE_\perp^*\right)^2=\left(\vect{\BB}-\vect{\EE}_\parallel\right)^2$ and $b$ is an scalar parameter controlling the magnitude of $\sigma_\perp$. Equations~(\ref{eq:sigmapar}) and~(\ref{eq:sigmaperp}) have a pair of analytic solutions:
\begin{align}
\begin{split}
\vect{\EE}_\parallel\left(t\right)=\:&\vect{\EE}_\parallel\left(0\right)\times e^{-\sigma_\parallel t}
\end{split}\label{eq:StrangCurrentPar}\\
\begin{split}
\vect{\EE}_\perp\hspace{-3pt}\left(t\right)=\:&\left[\EE_\perp^*\hspace{-3pt}\left(0\right)+\frac{\EE_\perp^*\hspace{-3pt}\left(0\right)\left[\EE_\perp\hspace{-3pt}\left(0\right)-\EE_\perp^*\hspace{-3pt}\left(0\right)\right]\times e^{-b\sigma_\parallel t}}{\EE_\perp\hspace{-3pt}\left(0\right)-\left[\EE_\perp\hspace{-3pt}\left(0\right)-\EE_\perp^*\hspace{-3pt}\left(0\right)\right]\times e^{-b\sigma_\parallel t}}\right]\\
&\times\frac{\vect{\EE}_\perp\hspace{-3pt}\left(0\right)}{\EE_\perp\hspace{-3pt}\left(0\right)}.
\end{split}\label{eq:StrangCurrentPerp}
\end{align}
During our numerical simulations), we usually choose $d=5.0$, and $b=0.1$, and solve equation~(\ref{eq:StrangCurrentPar}) prior to equation~(\ref{eq:StrangCurrentPerp}) in a Strang splitting scheme in direct analogy to the implementation described in section~\ref{sec:augmented_system}. This resistivity model ensures the validity of the force-free regime throughout time, in other words, the evolution is driven towards a force-free state
\begin{align}
\begin{split}
\vect{\EE}\cdot\vect{\BB}&\rightarrow0 \\
\vect{\BB}^2-\vect{\EE}^2&\rightarrow 0 \qquad\text{ : }\vect{\BB}^2<\vect{\EE}^2.
\end{split}
\end{align}

\section{Optical depth to resonant cyclotron scattering}
\label{sec:optical_thickness} 

For the presented modeling of the optical thickness of highly magnetised force-free plasmas around magnetars (see section~\ref{sec:emission_processes}), we adapt the techniques describing resonant scattering as presented by \citet{Beloborodov2013} (from now on Be13). In the following, we will give a short review of the underlying equations. In order to derive the optical thickness $\tau$, we integrate equation~(Be13/A15),
\begin{align}
\frac{\text{d}\tau}{\text{d} s}=2\pi^2 r_e\frac{c}{\omega}\frac{\xi}{\left|\tilde{\mu}\right|}n_e\left[f_e\left(p_1\right)+f_e\left(p_2\right)\right].
\label{eq:optical_thickness_integrand}
\end{align}
Here, $r_e=e^2/m_e c^2$ denotes the photon wavelength,
$\omega$ the frequency of the seed photon (we consider 1\,keV photons), and $\xi=1$ or $\xi=\tilde{\mu}^2$ depending on the photon polarisation ($\perp$ or $\parallel$, respectively). The relativistic particles require the specification of the quantities $\mu=\cos\vartheta$ and $\tilde{\mu}=\cos\tilde{\vartheta}$, where $\vartheta$ is the angle between the photon path and the magnetic field $\vect{B}$ in the lab frame and $\tilde{\vartheta}$ in the rest frame of the electron. The dimensionless momenta $p_{1,2}$ correspond to the electron (or positron) velocities favored by the resonant scattering model. As both polarisations yield similar results, we only consider the slightly dominant $\perp$ orientation for our model. \citet{Beloborodov2013} estimated that the contribution of non-resonant scattering to the optical depth is negligible and will not be considered in our calculations (see, however, footnote \ref{foot:one}).

Following Be13, we employ the so-called waterbag model as a distribution function for electron (or positron) momenta. In analogy to a two-fluid model, the distribution function is characterised by the two parameters (dimensionless momenta) $p_+$ and $p_-$, with the overall shape
\begin{align}
f_e\left(p\right)=\left\{\begin{array}{ccc}
(p_+ - p_-)^{-1}& : & p_-<p<p_+ \\ 
0 & : & \text{else}
\end{array}\right..
\label{eq:waterbag}
\end{align}
Applying the waterbag model (\ref{eq:waterbag}) in equation~(\ref{eq:optical_thickness_integrand}) selects the relevant electron (or positron) momenta for the scattering process. The distribution of this normalisation factor throughout the magnetosphere especially depends on the flow direction of charges along $\vect{B}$. As described in section 5.2 of Be13, we adjust their model according to a flow of electrons (or positrons) which turns back to the central object when field lines cross the equator. We apply this to all field lines crossing regions with $B<10^{13}$\,G (this holds everywhere except in the inner coronal region of strong closed magnetic field lines).

\label{lastpage}

\end{document}